\documentclass[11pt,a4paper]{article}
\usepackage{jheppub}
\pdfoutput=1
\usepackage{amsmath} 
\usepackage{graphicx} 
\usepackage{amsthm}
\usepackage{amssymb} 
\usepackage{hyperref}
\usepackage{srcltx}
\usepackage[justification=justified,font=small]{caption}
\usepackage{subfig}

\newcommand{\qbar}{\bar{q}}
\newcommand{\wbar}{\bar{\omega}}

\title{Bosonic excitations of the $AdS_4$ Reissner-Nordstr\"{o}m black hole}
\author{Richard A.~Davison and}
\author{Nikolaos K.~Kaplis}
\affiliation{Rudolf Peierls Centre for Theoretical Physics,\\ 1 Keble Road, Oxford OX1 3NP, United Kingdom}
\emailAdd{r.davison1@physics.ox.ac.uk}
\emailAdd{n.kaplis1@physics.ox.ac.uk}
\abstract{
We study the long-lived modes of the charge density and energy density correlators in the strongly-coupled, finite density field theory dual to the $AdS_4$ Reissner-Nordstr\"{o}m black hole. For small momenta $q\ll\mu$, these correlators contain a pole due to sound propagation, as well as a pole due to a long-lived, purely imaginary mode analogous to the $\mu=0$ hydrodynamic charge diffusion mode. As the temperature is raised in the range $T\lesssim\mu$, the sound attenuation shows no significant temperature dependence. When $T\gtrsim\mu$, it quickly approaches the $\mu=0$ hydrodynamic result where it decreases like $1/T$. It does not share any of the temperature-dependent properties of the `zero sound' of Landau Fermi liquids observed in the strongly-coupled D3/D7 field theory. For such small momenta, the energy density spectral function is dominated by the sound mode at all temperatures, whereas the charge density spectral function undergoes a crossover from being dominated by the sound mode at low temperatures to being dominated by the diffusion mode when $T\sim\mu^2/q$. This crossover occurs due to the changing residue at each pole. We also compute the momentum dependence of these spectral functions and their corresponding long-lived poles at fixed, low temperatures $T\ll\mu$.}

\begin{document}
\begin{flushright}OUTP-11-56P \end{flushright}
\maketitle
\flushbottom

\section{Introduction}
\paragraph{}
The AdS/CFT, or gauge/gravity, correspondence \cite{Maldacena:1997re,Gubser:1998bc,Witten:1998qj,Aharony:1999ti} is a very useful tool which allows one to study the properties of certain strongly-coupled field theories via their dual gravitational theories. In particular, the thermodynamic and near-equilibrium properties of such strongly-coupled field theories can be obtained relatively easily from their dual gravitational descriptions. Initial studies of these properties concentrated on field theories at non-zero temperature $T$ (most notably $\mathcal{N}=4$ $SU(N_c)$ supersymmetric Yang-Mills theory with $N_c\rightarrow\infty$) and were motivated by experimentally-observed properties of thermal field theories \cite{CasalderreySolana:2011us}. For perturbations whose frequency $\omega$ and momentum $q$ are much less than $T$, these field theories were found to obey the laws of hydrodynamics and their transport coefficients such as shear viscosity, charge diffusion constant etc. were calculated (see \cite{Policastro:2002se,Policastro:2002tn,Bhattacharyya:2008jc} and subsequent work). More recently, there has been a lot of interest in studying field theories at \textit{zero} temperature but with a non-zero density of a conserved global $U(1)$ charge - these are analogues of strongly-coupled condensed matter systems with a non-zero density of particles (see \cite{Hartnoll:2009sz,Herzog:2009xv,McGreevy:2009xe,Hartnoll:2011fn} for some introductions to the field). 

\paragraph{}
There have been numerous studies of the excitations present when $T=0$ and $\omega,q\ll\mu$ in specific strongly-coupled field theories, and one common feature amongst many examples is a propagating longitudinal mode with the dispersion relation 
\begin{equation}
\label{eq:sounddispersion}
\wbar=\pm v_s\qbar-i\Gamma_0\qbar^2 +O\left(\qbar^3\right),
\end{equation}
where 
\begin{equation}
\wbar=\frac{\omega}{\mu}, \;\;\;\;\;\; \text{and} \;\;\;\;\;\;  \qbar=\frac{q}{\mu}.
\end{equation}
This is a feature of probe brane theories in different dimensions and with different UV symmetries (where the conserved charge is a density of fundamental matter - at least in the case where the background is derived from string theory) \cite{Karch:2008fa,Nickel:2010pr,Kulaxizi:2008kv,Ammon:2011hz,Kulaxizi:2008jx,Hung:2009qk,HoyosBadajoz:2010kd,Lee:2010ez,Bergman:2011rf} and also the 4D bulk Einstein-Maxwell theory with a cosmological constant (where the conserved charge is an R-charge density) \cite{Edalati:2010pn}.

\begin{figure*}
\begin{center}
\includegraphics[scale=1.0]{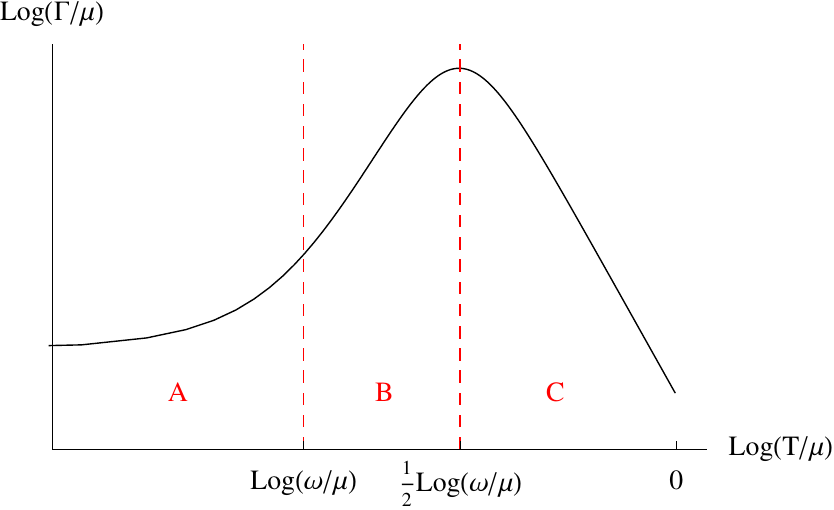}
\caption{The sound attenuation $\Gamma$ in an LFL as a function of temperature, at fixed $\omega$ and $\mu$. A is the collisionless quantum regime, B is the collisionless thermal regime and C is the hydrodynamic regime.}
\label{fig:LFLplot}
\end{center}
\end{figure*}

\paragraph{}
In one of these probe brane theories (the D3/D7 theory in (3+1) dimensions), the behaviour of this zero temperature sound mode was analysed as the temperature was increased from $T=0$ and it was found to behave similarly to the `zero sound' mode due to the oscillation of the Fermi surface of a Landau Fermi liquid (LFL), despite the fact that some of its other properties are quite different from an LFL (e.g. its heat capacity is proportional to $T^6$ rather than $T$) \cite{Davison:2011ek}. The theory of an LFL is valid when $T\ll\mu$ and $\omega,q\ll\mu$ and it predicts three different regimes for the behaviour of the sound attenuation, as shown in figure \ref{fig:LFLplot} \cite{LFL1,LFL2,LFL3,LFL4,LFL5}.
At $T=0$, the sound mode has a non-zero attenuation $\Gamma$ proportional to $\omega^2/\mu$. As the temperature is increased from zero (with $\omega$ and $\mu$ fixed), the attenuation of the sound mode remains approximately constant until $T/\mu\sim\omega/\mu$ (the `collisionless quantum regime' denoted by A in figure \ref{fig:LFLplot}). Above this, thermal excitations around the Fermi surface must be taken into account and collisions between these thermally-excited quasiparticles cause the sound attenuation to increase at a rate proportional to $T^2/\mu$ (the `collisionless thermal regime' B). When $T$ is sufficiently high such that $T/\mu\sim\sqrt{\omega/\mu}$, these thermal collisions become so frequent that the zero sound mode is no longer a long-lived mode. However, the thermal collisions support the hydrodynamic modes of sound and diffusion. This results in the `collisionless/hydrodynamic crossover' and the sound attenuation begins to decrease at a rate proportional to $\mu\left(\omega/T\right)^{2}$ as the hydrodynamic sound mode stabilises.

\paragraph{}
These features were reproduced precisely in the strongly-coupled D3/D7 field theory except that there was no sound propagation in the hydrodynamic regime - this can be explained by the fact that the fluctuations of the bulk metric, which generate the hydrodynamic sound mode of the dual field theory, were explicitly suppressed for consistency with the probe limit. The collisionless/hydrodynamic crossover is most clearly exhibited in the D3/D7 theory via the poles of the charge density Green's function in the complex frequency plane \cite{Davison:2011ek}. As the temperature is raised, the low temperature poles corresponding to sound propagation approach the imaginary axis and collide to form two purely imaginary poles, one of which becomes the hydrodynamic diffusion mode as the temperature is raised further. The crossover can also be seen by examining the charge density spectral function - the tall, narrow peak corresponding to the sound mode becomes smaller and moves towards the origin as the temperature is raised, eventually forming a diffusive peak around $\wbar=0$. A natural question to ask is whether similar behaviour is observed in the low temperature sound modes of other holographic theories at large chemical potential and hence whether they can all be characterised in this way as LFL-like `zero sound' modes or not.

\paragraph{}
In this paper, we study the strongly-coupled field theory dual to the RN-$AdS_4$ black hole solution of 4D Einstein-Maxwell theory with a cosmological constant. Many properties of this theory have been investigated in recent years as it is relatively simple and yet has very interesting behaviour. When $T=0$, the low energy behaviour of the field theory is governed by a $\text{CFT}_1$ dual to the $AdS_2$ factor of the black hole's $AdS_2\times \mathbb{R}^2$ near-horizon geometry \cite{Faulkner:2009wj}. If one considers a probe Dirac action for fermions in this background, the field theory operators dual to these fermions exhibit Fermi surfaces of a non-Fermi liquid type \cite{Cubrovic:2009ye,Liu:2009dm,Faulkner:2009wj}. If instead one considers a fermionic action which is the supersymmetric completion of the Einstein-Maxwell action, no such Fermi surfaces are observed \cite{Belliard:2011qq,Gauntlett:2011mf,Gauntlett:2011wm}. 

\paragraph{}
The quasinormal modes of the bulk bosonic fields, which correspond to the poles of the retarded Green's functions of the field theory energy-momentum tensor $T^{\mu\nu}$ and U(1) current $J^\mu$, have also been studied at zero temperature. At $T=0$ the transverse sector (i.e. transport perpendicular to the direction of the momentum flow) contains a branch cut along the negative imaginary frequency axis, and no long-lived modes (by which we mean no modes satisfying $\wbar\rightarrow0$ as $\qbar\rightarrow0$). At very small temperatures $T\ll\mu$, this branch cut becomes a series of poles, one of which has a dispersion relation of the form $\wbar=-i\mathcal{D}\qbar^2+O\left(\qbar^3\right)$ \cite{Edalati:2010hk,Brattan:2010pq,Ge:2010yc}. This is an analogue of the well-known hydrodynamic shear diffusive mode of the energy-momentum tensor \cite{Policastro:2002se}. At $T=0$, the longitudinal sector (i.e. transport parallel to the direction of the momentum flow) also contains a branch cut along the negative imaginary frequency axis. In addition to this, it has two propagating modes with dispersion relations of the form (\ref{eq:sounddispersion}) with $v_s=\sqrt{(\partial P/\partial\epsilon)}|_{T=0}$ and $\Gamma_0\approx\mu\eta_0/2(\epsilon+P)$, where $\eta_0$ is the `zero temperature viscosity' derived via the `Kubo formula' 
\begin{equation}
\eta_0=-\lim_{\omega\rightarrow0}\frac{1}{\omega}\text{Im}\left[G^R_{T^{xy}T^{xy}}\left(\omega,0\right)\right]\bigg|_{T=0},
\end{equation}
$\epsilon$ and $P$ are the field theory's energy density and pressure respectively, and the `$\approx$' signifies that these are equal to within about $10\%$ \cite{Edalati:2009bi,Edalati:2010pn}.\footnote{The sound attenuation $\Gamma_0$ is only known numerically and thus an exact comparison cannot be made.} A priori, this result is not obvious. It suggests that when $T=0$, $\mu$ acts as an `effective hydrodynamic scale', at least as far as sound propagation is concerned.

\paragraph{}
In this paper, we first study numerically the behaviour of the longitudinal poles for a fixed momentum $\qbar<1$ as the temperature is increased from $T=0$ to $T\gg\mu$. Of particular interest to us is the behaviour of the sound mode and whether there is a collisionless/hydrodynamic crossover as in the D3/D7 theory and Landau's theory of Fermi liquids. We find that the attenuation of the sound mode shows no significant temperature dependence over the range $T\lesssim\mu$ where we may have expected to find LFL-like behaviour and above this it approaches the $\mu=0$ hydrodynamic result \cite{Herzog:2003ke,Miranda:2008vb} where the attenuation decreases like $T^{-1}$. Its speed is approximately $1/\sqrt{2}$ at all temperatures. This is in complete contrast to the results for an LFL (as shown in figure \ref{fig:LFLplot}). Such an outcome is not particularly surprising since all available evidence suggests that this field theory is \textit{not} an LFL. However, this is also true of the D3/D7 theory and yet it possesses an LFL-like zero sound mode. Our results show that this kind of mode is \textit{not} generic to all strongly-coupled field theories at large chemical potential which have a gravitational dual.

\paragraph{}The other long-lived, longitudinal mode of the theory corresponds to a purely imaginary pole that forms when the branch cut along the negative imaginary frequency axis dissolves at non-zero temperatures. This mode becomes more stable as the temperature increases and when $T\gg\mu$ its dispersion relation is that of the $\mu=0$ hydrodynamic charge diffusion mode \cite{Herzog:2002fn,Miranda:2008vb}.

\paragraph{}
In addition to the poles of the Green's functions, we also compute the spectral functions of energy density and charge density as the temperature is increased at fixed $\qbar<1$. The energy density spectral function is dominated by the peak corresponding to sound propagation at all temperatures. In contrast to this, the charge density spectral function undergoes a crossover from being dominated by the sound peak at low temperatures to being dominated by the diffusion peak at high temperatures $T\gg\mu^2/q$. Note that this mechanism is quite different than in the D3/D7 theory where the sound poles collide to form the diffusion pole. Here, the sound and diffusion poles coexist at all non-zero temperatures (that we can access numerically) but their residues change and this results in the crossover. In the D3/D7 theory this crossover was reminiscent of that in an LFL, but here we know of no such comparison - in particular, the RN-$AdS_4$ crossover occurs outside of the `quantum liquid' range $T\ll\mu$ where we may expect an LFL-like theory to apply.

\paragraph{}
As well as the previous results regarding the temperature dependence of the poles and spectral functions at fixed momentum, we have also calculated their momentum dependence at various fixed temperatures. We find that the sound and purely imaginary modes exist at all non-zero temperatures that we can access numerically ($T\ge0.0219\mu$) with the dispersion relations
\begin{equation}
\label{eq:sounddispersionrelationallT}
\wbar=\frac{\qbar}{\sqrt{2}}-i\Gamma(T/\mu)\qbar^2+O(\qbar^3),
\end{equation}
and
\begin{equation}
\label{eq:diffusiondispersionrelationallT}
\wbar=-iD(T/\mu)\qbar^2+O(\qbar^3),
\end{equation}
and we compute the functions $D(T/\mu)$ and $\Gamma(T/\mu)$ numerically. For this reason, and the fact that it becomes the $\mu=0$ hydrodynamic charge diffusion mode in the $T\gg\mu$ limit, we will refer to this purely imaginary mode as the diffusion mode. When $\qbar\gg1$, these modes no longer dominate the low-energy properties of the theory and we must consider additional poles of the Green's functions also.

\paragraph{}
Finally, we investigate the properties of sound propagation in the theory, at some fixed momentum $q$, over the $\left(T,\mu\right)$ plane. As discussed above, it is known that when $\mu=0$, $T$ is the hydrodynamic scale and sound will propagate provided its momentum satisfies $q\ll T$. We also know that when $T=0$, there is an effective hydrodynamic scale $\mu$ in that there is a long-lived sound mode provided that $q\ll\mu$. We find that when both of these scales are non-zero, there will be a long-lived sound mode provided that any \textit{one} of them is much larger than the momentum.  In other words, sound will propagate for \textit{any} value of $\mu/T$, provided that one considers small enough momenta.

\paragraph{}
The structure of the paper is as follows. In section \ref{sec:backgroundandfluctuations} we present the RN-$AdS_4$ solution of the four dimensional Einstein-Maxwell theory with a cosmological constant, as well as the gauge-invariant fluctuations of the bulk fields that we use to compute the poles of the retarded Green's functions and the spectral functions. Section \ref{sec:Tdependencesound} contains the numerical results showing how the properties of the sound mode change as we increase the temperature at fixed $\qbar<1$. In particular, we demonstrate that these properties are significantly different from those of an LFL and of the D3/D7 theory. In section \ref{sec:furtherTdependentproperties}, we explore how the other long-lived mode of the system - the diffusion mode - behaves as $T$ is increased for $\qbar<1$. We also examine how the energy density and charge density spectral functions vary with $T$ for $\qbar<1$. Section \ref{sec:qdependenceresults} contains results for the poles and spectral functions as a function of $\qbar$ at fixed temperature $T<\mu$. We then study the existence of an effective hydrodynamic scale in section \ref{sec:effectivehydroscale} by examining the properties of the sound mode as a function of both $q/\mu$ and $q/T$. This is followed by our conclusions and a discussion of the behaviour of the sound mode compared to that of the D3/D7 theory in section \ref{sec:discussion}. Appendix \ref{sec:appendix} contains the equations of motion and off-shell action for the aforementioned gauge-invariant fluctuations.

\section{The RN-$AdS_4$ background and fluctuations}
\label{sec:backgroundandfluctuations}
%
\subsection{The action and background solution}
\paragraph{}
The gravitational theory we study is the four dimensional Einstein-Maxwell theory with a cosmological constant, which has the action
\begin{equation}
\label{eq:einsteinmaxwellaction}
S=\frac{1}{2\kappa_4^2}\left[\int_\mathcal{M} d^4x\sqrt{-g}\left(R-2\Lambda-L^2F_{\mu\nu}F^{\mu\nu}\right)+2\int_{\partial\mathcal{M}}d^3x\sqrt{|h|}\left(\mathcal{K}+\text{counterterms}\right)\right],
\end{equation}
where $\Lambda=-3/L^2$, $h$ is the induced metric on the boundary of the spacetime, $\mathcal{K}$ is the extrinsic curvature on this boundary and $F_{\mu\nu}$ is the field strength of a $U(1)$ gauge field $A_\mu$. This is a consistent truncation of $D=11$ supergravity \cite{Chamblin:1999tk,Cvetic:1999xp}. The counterterms are discussed in the appendix.

\paragraph{}
This theory has a charged, asymptotically-AdS black hole solution with a planar horizon: the planar $AdS_4$ Reissner-Nordstr\"{o}m black hole (RN-$AdS_4$)
\begin{equation}
\label{eq:backgroundsolution}
\begin{aligned}
ds_{4}^2&=-\frac{r^2f(r)}{L^2}dt^2+\frac{r^2}{L^2}dx^2+\frac{r^2}{L^2}dy^2+\frac{L^2}{r^2f(r)}dr^2,\\
f(r)&=1-(1+Q^2)\left(\frac{r_0}{r}\right)^3+Q^2\left(\frac{r_0}{r}\right)^4,\\
A_t&=\frac{Qr_0}{L^2}\left(1-\frac{r_0}{r}\right),
\end{aligned}
\end{equation} 
where $r$ is the bulk radial co-ordinate, $r_0$ is the position of the horizon and $Q$ determines the $U(1)$ charge of the black hole. In the full $D=11$ supergravity theory, it is the decoupling limit of the geometry created by a stack of M2-branes which are rotating (in a specific way) in the directions transverse to their worldvolume \cite{Cvetic:1999xp}. The bulk $U(1)$ gauge field is dual to a $U(1)$ R-current in the field theory on this worldvolume.

\paragraph{}
This solution has one tunable dimensionless parameter $Q$ which determines the ratio of the chemical potential in the field theory to the temperature
\begin{equation}
\begin{aligned}
\frac{\mu}{T}=\frac{4\pi Q}{3-Q^2}.
\end{aligned}
\end{equation}
It takes values between $0$ (the zero chemical potential limit) and $\sqrt{3}$ (the zero temperature limit). The thermodynamics of the dual field theory are well-known \cite{Chamblin:1999tk}. In particular, we note that the entropy density has the unusual property of being non-zero when $T=0$.
\subsection{Fluctuations around equilibrium}
\paragraph{}
We are interested in the response of the field theory to small perturbations around the equilibrium state - this is encoded holographically in the linear response of the black hole to perturbations around the background solution (\ref{eq:backgroundsolution}):
\begin{equation}
\begin{aligned}
g_{\mu\nu}&\rightarrow g_{\mu\nu}+h_{\mu\nu},\\
A_\mu&\rightarrow A_\mu+a_\mu.
\end{aligned}
\end{equation}
We use the rotational invariance in the $(x,y)$-plane to choose the momentum to flow only in the $x$-direction of the field theory, without loss of generality. We may then classify fluctuations according to their parity under $y\rightarrow-y$. The fluctuations which are even under this operation ($h_{xx},\;h_{yy},\;h_{rr},\;h_{tt},\;h_{rt},\;h_{rx},\;h_{xt},\;a_r,\;a_t\text{ and }a_x$) decouple from those which are odd ($h_{yr},\;h_{yx},\;h_{yt}\text{ and }a_y$) at linear order \cite{Herzog:2002fn}. The indices are raised and lowered with the background metric. In the following we are interested only in the even fluctuations which we refer to as `longitudinal' henceforth, as they encode the response of the fields parallel to the direction of momentum flow. The metric and gauge field fluctuations are coupled within this longitudinal sector which tells us that the retarded Green's functions of the longitudinal components of the field theory energy-momentum tensor $T^{\mu\nu}$ and U(1) conserved current $J^{\mu}$ are not independent. 

\paragraph{}
We are particularly interested in two properties of the retarded Green's functions $G^R_{\mathcal{O}_A\mathcal{O}_B}$. The first is the poles of the Green's functions in the complex frequency plane. These poles correspond to the field theory excitations - the real part of each pole is its propagating frequency and the imaginary part is its decay rate. We are primarily interested in the long-lived excitations - those with the smallest imaginary part. Note that if any excited bulk fields are coupled, their dual field theory operators share a common set of Green's function poles.

\paragraph{}
The second object of interest to us is the matrix of spectral functions
\begin{equation}
\chi_{AB}\left(\omega,q\right)\equiv i\left(G^R_{\mathcal{O}_A\mathcal{O}_B}\left(\omega,q\right)-G^{R}_{\mathcal{O}_B\mathcal{O}_A}\left(\omega,q\right)^*\right),
\end{equation}
which tells us the rate of work done on the system by small external sources for $\mathcal{O}_A$ and $\mathcal{O}_B$ with frequency $\omega$ (see, for example, \cite{Hartnoll:2009sz}). Modes which couple strongly to external sources in this way are visible in the spectral functions as tall, narrow peaks centred on the propagating frequency and with a width proportional to their decay rate. Such a peak will be produced by a pole of the Green's function with small imaginary part provided that the residue of the Green's function is large enough at this pole and that there are no other poles near it in the complex frequency plane. Unlike the existence of a pole, the residue at a pole differs between the Green's functions of a set of coupled operators and hence despite the fact that they have a shared set of Green's function poles, the spectral functions of coupled operators can be very different. We are interested in the energy density spectral function $\chi_{T^{tt}T^{tt}}\equiv\chi_{\epsilon\epsilon}$ and the charge density spectral function $\chi_{J^tJ^t}\equiv\chi_{QQ}$, which are real quantities, and tell us which modes couple strongly to external sources of energy density and charge density respectively.
\subsection{Gauge-invariant variables and Ward identities}
\paragraph{}
The retarded Green's functions can be computed from the on-shell action of the gravitational theory which generically has the form
\begin{equation}
S_{\text{on-shell}}=\int_{r\rightarrow\infty, \omega>0}\frac{d\omega d^2q}{\left(2\pi\right)^3}\Bigl[\phi_{I}(r,-\omega,-q)\mathcal{G}_{IJ}\partial_r\phi_{J}(r,\omega,q)+\phi_{I}(r,-\omega,-q)\mathcal{C}_{IJ}\phi_{J}(r,\omega,q)\Bigr],
\end{equation}
where $\phi_{I}$ label the perturbations of bulk fields which are dual to field theory operators $\mathcal{O}_I$. To obtain the retarded Green's function of two operators $G^R_{\mathcal{O}_A\mathcal{O}_B}\left(\omega,q\right)$, one must find solutions to the bulk field equations which satisfy ingoing conditions at the black hole horizon and asymptote to
\begin{equation}
\phi_I\left(r\rightarrow\infty,-\omega,-q\right)\rightarrow\begin{cases} 1, & I=A \\ 0, & I\ne A\end{cases} \hspace{7mm} \text{and} \hspace{7mm}\phi_J\left(r\rightarrow\infty,\omega,q\right)\rightarrow\begin{cases} 1, & J=B \\ 0, & J\ne B\end{cases},
\end{equation}
near the boundary. $G^R_{\mathcal{O}_A\mathcal{O}_B}\left(\omega,q\right)$ is then given by the integrand of the on-shell action evaluated on these solutions \cite{Son:2002sd,Kaminski:2009dh}.

\paragraph{}
The longitudinal sector of our theory contains ten fields whose excitations are coupled. The equations of motion and on-shell action for the excited longitudinal fields can be simplified considerably by noting that the theory has a U(1) gauge symmetry which acts on the gauge field fluctuations as
\begin{equation}
a_{\mu}\rightarrow a_{\mu}-\partial_\mu\Lambda,
\end{equation}
and a diffeomorphism symmetry which acts as 
\begin{equation}
\begin{aligned}
\label{eq:diffeomorphisms}
&h_{\mu\nu}\rightarrow h_{\mu\nu}-\nabla_\mu\xi_\nu -\nabla_\nu\xi_\mu, \\
&a_{\mu}\rightarrow a_{\mu}-\xi^\alpha\nabla_\alpha A_\mu-A_\alpha\nabla_{\mu}\xi^\alpha, 
\end{aligned}
\end{equation}
to linear order, where $\nabla$ is the covariant derivative with respect to the background metric \cite{waldtextbook}. We can form two linearly-independent variables which are invariant under these transformations
\begin{equation}
\begin{aligned}
Z_1(r,\omega,q)&=\omega a_x(r,\omega,q)+ qa_t(r,\omega,q)-\frac{qL^2A_t'(r)}{2r}h_{yy},\\
Z_2(r,\omega,q)&=\frac{2\omega q}{r^2}h_{xt}+\frac{\omega^2}{r^2}h_{xx}+\frac{q^2}{r^2}h_{tt}+\frac{q^2f(r)}{r^2}h_{yy}\left(1+\frac{rf'(r)}{2f(r)}-\frac{\omega^2}{q^2f(r)}\right),
\end{aligned}
\end{equation}
and to linear order in the fluctuations, we can write our theory in terms of these variables. In particular, this reduces the set of ten coupled equations to a set of two. It also allows us to write the on-shell action in the form
\begin{equation}
S_{\text{on-shell}}=\int_{r\rightarrow\infty, \omega>0} \frac{d\omega d^2q}{\left(2\pi\right)^3}\Bigl[Z_i\left(r,-\omega,-q\right)\mathcal{G}_{ij}\partial_rZ_j\left(r,\omega,q\right)+\phi_{I}\left(r,-\omega,-q\right)\mathcal{C}_{IJ}\phi_{J}\left(r,\omega,q\right)\Bigr],
\end{equation}
where $i=1,2$ and $\phi_{I}$ denote the fundamental fluctuations $\left\{h_{tt}, h_{xx}, h_{yy}, h_{tx},a_t,a_x\right\}$. The $\mathcal{C}_{IJ}$ terms are analytic in $\omega,q$ and hence contribute only contact terms to the retarded Green's functions (i.e. terms analytic in $\omega,q$). These $Z_i$ variables are a generalisation of those of \cite{Kovtun:2005ev} to non-zero chemical potential (and in 3+1, rather than 4+1, dimensions).

\paragraph{}
Written in this way we see explicitly that, neglecting contact terms, the on-shell action for a solution that has the form $a_t\left(r\rightarrow\infty,\pm\omega,\pm q\right)\rightarrow1$ (with all others fields zero in this limit) differs from that for a solution with $a_x\left(r\rightarrow\infty,\pm\omega,\pm q\right)\rightarrow1$ (with all other fields zero in this limit) only by the factor $q^2/\omega^2$ in the definition of the variables $Z_i$, and similarly for the other fields.

\paragraph{}
This property of bulk gauge-invariance thus generates a number of relationships between the retarded Green's functions of the corresponding field theory operators:
\begin{equation}
G^R_{J^xJ^t}=\frac{\omega}{q}G^R_{J^tJ^t}, \hspace{10mm} G^R_{J^xJ^x}=\frac{\omega^2}{q^2}G^R_{J^tJ^t},
\end{equation}
\begin{equation}
\begin{aligned}
&G^R_{T^{xx}J^t}=\frac{\omega^2}{q^2}G^R_{T^{tt}J^t}, \hspace{5mm} G^R_{T^{tx}J^t}=\frac{\omega}{q}G^R_{T^{tt}J^t}, \hspace{5mm} G^R_{T^{yy}J^t}=\left(1-\frac{\omega^2}{q^2}\right)G^R_{T^{tt}J^t}, \hspace{2mm} \ldots,
\end{aligned}
\end{equation}
\begin{equation}
\begin{aligned}
G^R_{T^{xx}T^{tt}}=\frac{\omega^2}{q^2}G^R_{T^{tt}T^{tt}}, \hspace{5mm} G^R_{T^{tx}T^{tt}}=\frac{\omega}{q}G^R_{T^{tt}T^{tt}}, \hspace{2mm} \ldots,
\end{aligned}
\end{equation}
where the `$\ldots$' represents other similar relations, and these equations should be understood to hold up to contact terms. These are precisely the Ward identities of the field theory.

\paragraph{}
Thus not only do these gauge-invariant variables simplify the equations of motion for the bulk fluctuations, they also explicitly encode the Ward identities of the field theory. The contribution of the contact terms to the diagonal retarded Green's functions is purely real, and thus they don't affect our results for the spectral functions $\chi_{\epsilon\epsilon}$ and $\chi_{QQ}$. Note that the contact terms cannot be written in terms of these gauge-invariant variables - we believe that this is because the linear diffeomorphism transformations (\ref{eq:diffeomorphisms}) are not the correct ones to apply to the action which is quadratic in fluctuations.

\paragraph{}We note that these are not the only possible gauge-invariant choice of variables. Another choice is the Kodama-Ishibashi variables which involve radial derivatives of the bulk fields, and have the advantage that the two equations of motion in these variables decouple \cite{Kodama:2003kk}.

\subsection{Equations of motion and on-shell action in dimensionless variables}
\paragraph{}
It is convenient to work with the dimensionless radial co-ordinate $u\equiv r/r_0$. For $T>0$, we use the gauge-invariant variables
\begin{equation}
\begin{aligned}
&\bar{Z}_1\left(u,\omega,q\right)=\wbar a_x+\qbar a_t-\frac{\qbar\mu}{2u}h^y_y,\\
&\bar{Z}_2\left(u,\omega,q\right)=i\mu\left[2\wbar\qbar h^x_t+\wbar^2h^x_x-\qbar^2f(u)h^t_t+\qbar^2f(u)\left(1+\frac{uf'(u)}{2f(u)}-\frac{\wbar^2}{\qbar^2f(u)}\right)h^y_y\right],
\end{aligned}
\end{equation}
where $f(u)=1-(1+Q^2)/u^3+Q^2/u^4$. The equations of motion and action in these variables are given in the appendix. To compute the poles and spectral functions, we used the numerical procedure described in \cite{Kaminski:2009dh}.

\paragraph{}
At $T=0$, we use the Kodama-Ishibashi variables and follow the methods described in \cite{Edalati:2010pn}. The equations of motion in these variables are given in appendix A of \cite{Edalati:2010pn}. We could only obtain accurate numerics at $T=0$ above $\qbar\gtrsim0.1$, and hence we only show $T=0$ results in this range.
\section{Temperature dependence of the sound mode}
\label{sec:Tdependencesound}
\paragraph{}
Our primary motivation for studying this theory is that it supports stable, propagating excitations of energy and charge density at zero temperature and large chemical potential $\qbar\ll1$. These sound modes at zero temperature have a dispersion relation of the form (\ref{eq:sounddispersion}) where the speed is $v_s=1/\sqrt{2}\;$ \cite{Edalati:2010pn}. We want to know what effect the increase of temperature has upon this mode - in particular we are looking to see if it shares the characteristics of the `zero sound' mode of a Landau Fermi liquid. This comparison can be made by studying the sound attenuation as a function of temperature for $T\ll\mu$ and looking for the three different regimes shown in figure \ref{fig:LFLplot}.

\paragraph{}
Note that when $\mu=0$ and $\omega,q\ll T$, there are sound modes with the dispersion relation
\begin{equation}
\label{eq:herzogsound}
\omega=\pm\frac{1}{\sqrt{2}}q-i\frac{1}{8\pi T}q^2+O(q^3).
\end{equation}
At non-zero $\mu$, we would expect to recover these results when $\mu\ll\omega,q\ll T$, which is outside of the `quantum liquid' regime $T\ll\mu$ where any LFL-like behaviour would be present.

\begin{figure*}
\begin{center}
\includegraphics[scale=0.88]{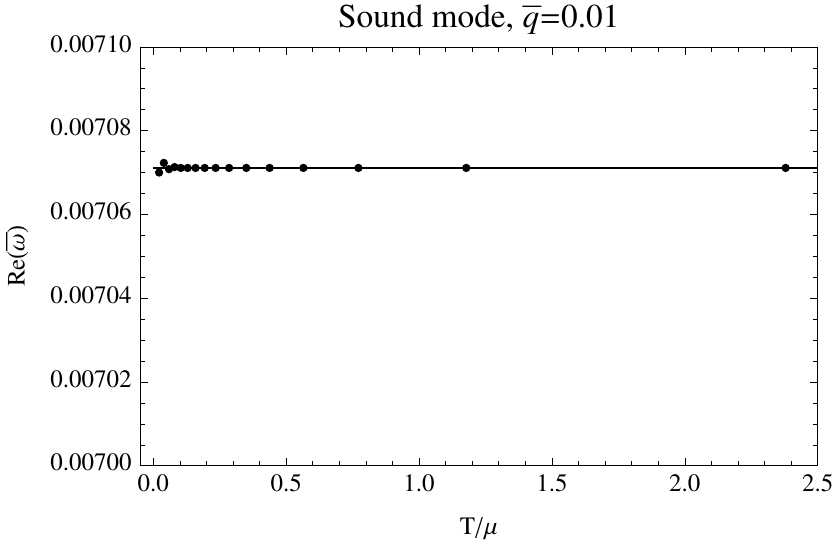}
\includegraphics[scale=0.88]{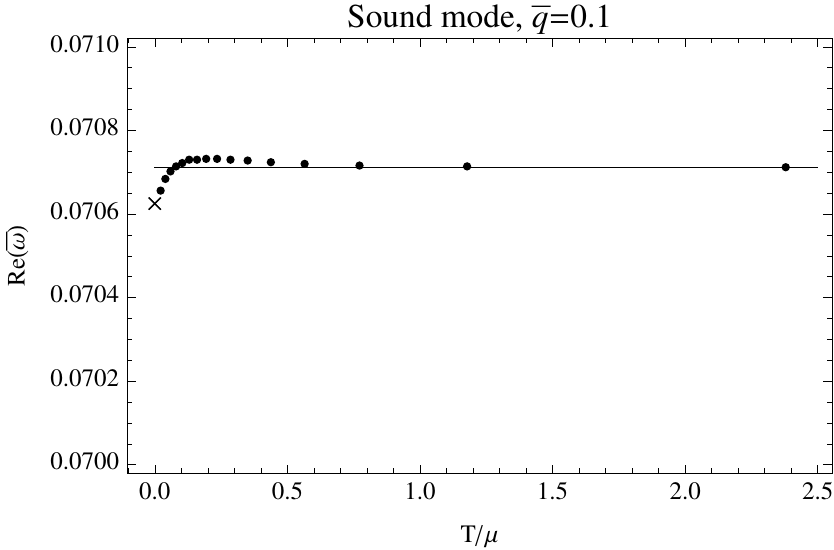}
\includegraphics[scale=0.88]{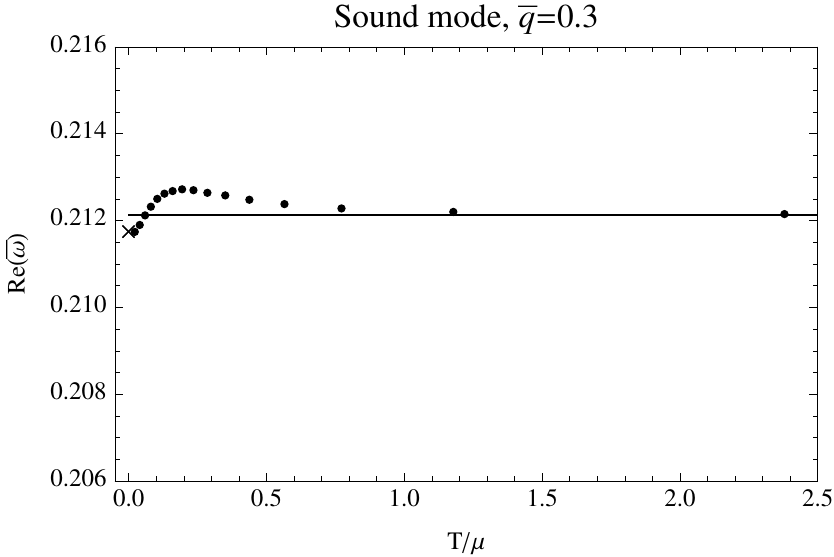}
\includegraphics[scale=0.88]{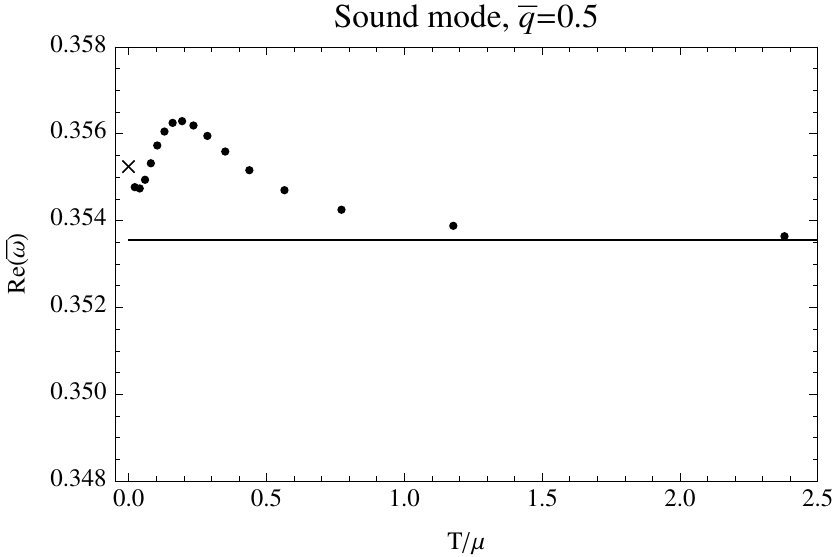}
\caption{Variation of the real part of the sound mode as the temperature is increased. The crosses mark the $T=0$ numerical results, the dots are the numerical results for $T>0$, and the solid lines are the $\mu=0$ analytic result (\ref{eq:herzogsound}).}
\label{fig:Tdependencerealsound}
\end{center}
\end{figure*}

\begin{figure*}
\begin{center}
\includegraphics[scale=0.88]{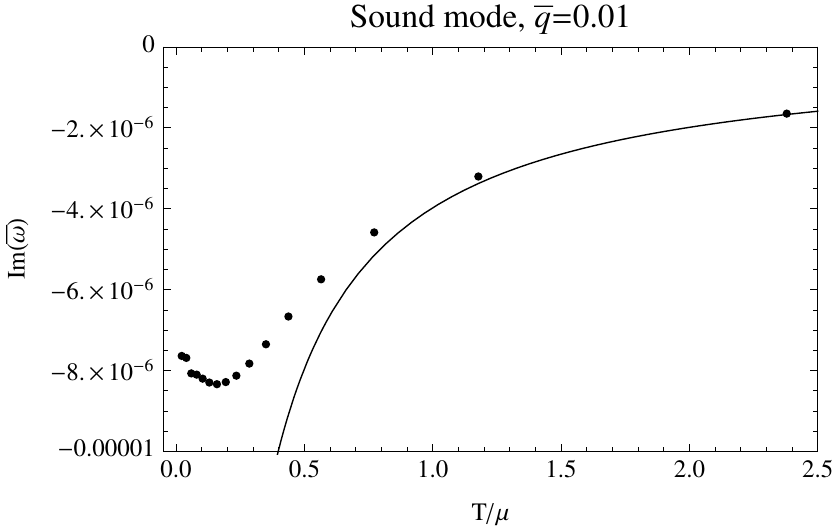}
\includegraphics[scale=0.88]{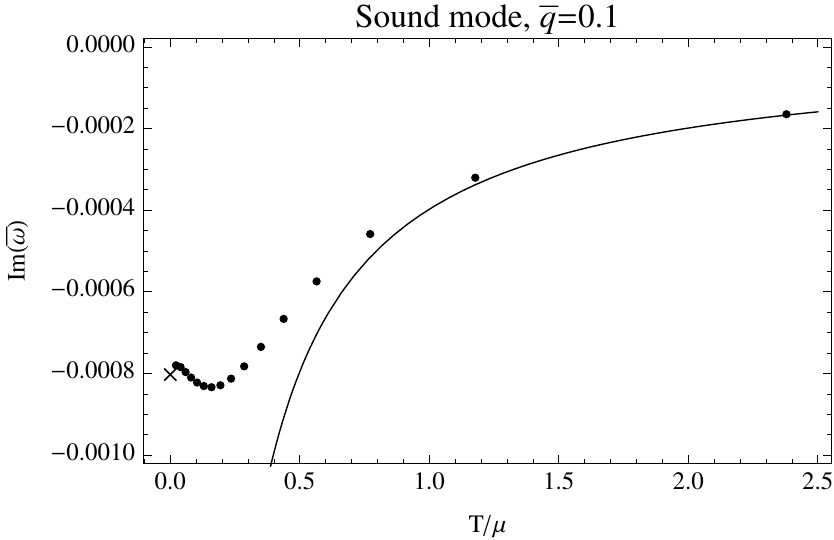}
\includegraphics[scale=0.88]{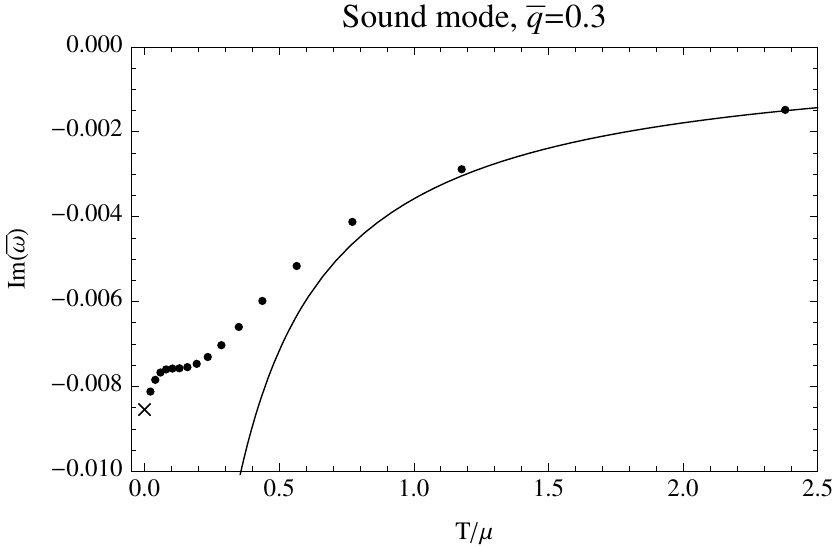}
\includegraphics[scale=0.88]{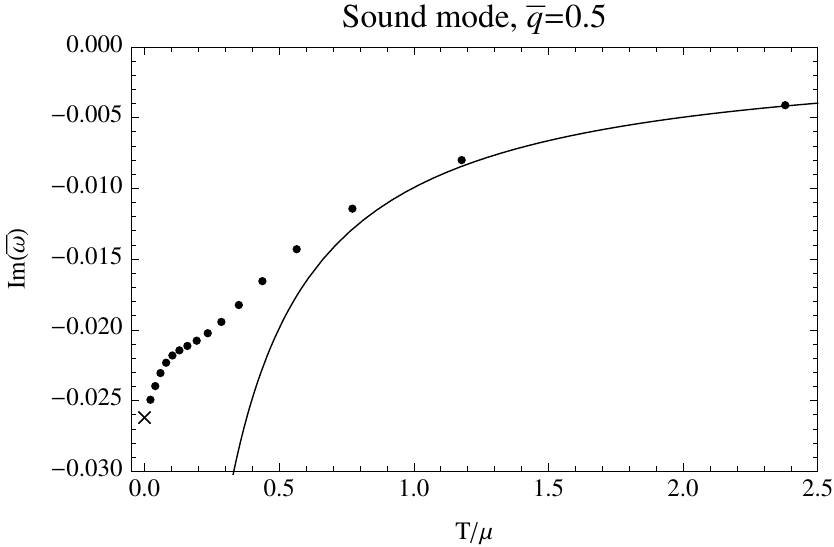}
\caption{Variation of the imaginary part of the sound mode as the temperature is increased. The crosses marks the $T=0$ numerical results, the dots are the numerical results for $T>0$, and the solid lines are the $\mu=0$ analytic result (\ref{eq:herzogsound}).}
\label{fig:Tdependenceimaginarysound}
\end{center}
\end{figure*}

\paragraph{}
The temperature dependence of the real and imaginary parts of the sound mode are shown in figures \ref{fig:Tdependencerealsound} and \ref{fig:Tdependenceimaginarysound} for various $\qbar<1$. Our finite temperature numerical results are shown along with $T=0$ numerical results (for $\qbar\ge0.1$ where we can obtain accurate results) and the $\mu=0$ analytic result (\ref{eq:herzogsound}).

\paragraph{}
The plots show that both the real and imaginary parts of the mode have a non-trivial temperature dependence. As the temperature is increased from zero, finite temperature corrections cause small changes to the real part of the sound mode whose sign depends upon the value of $\qbar$. At sufficiently high temperature, $T/\mu\gtrsim1$, the real part quickly asymptotes to the $\mu=0$ hydrodynamic result (\ref{eq:herzogsound}). The imaginary part of the sound mode shows similar behaviour. This is slightly surprising - it indicates that the $\mu=0$ result (\ref{eq:herzogsound}) is valid when $q\ll\mu\ll T$. We will comment upon this again in section \ref{sec:effectivehydroscale}. 

\begin{figure*}
\begin{center}
\includegraphics[scale=0.88]{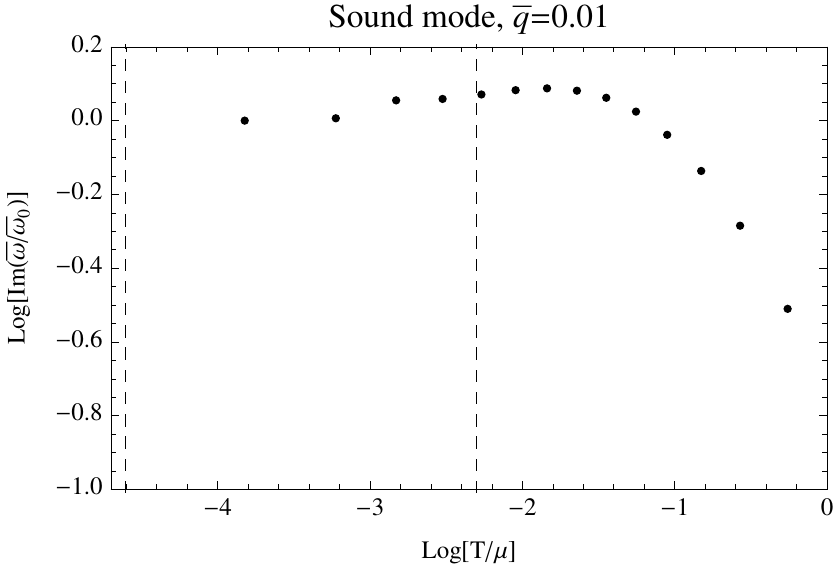}
\includegraphics[scale=0.88]{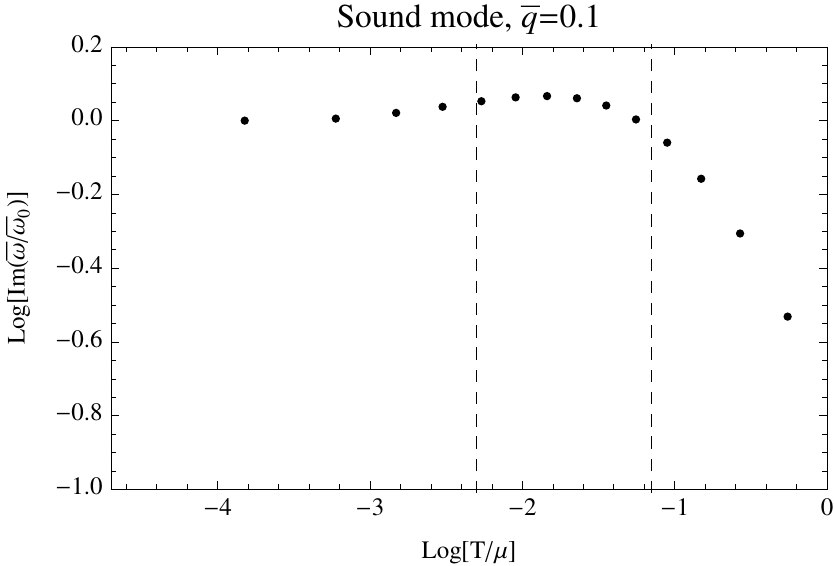}
\includegraphics[scale=0.88]{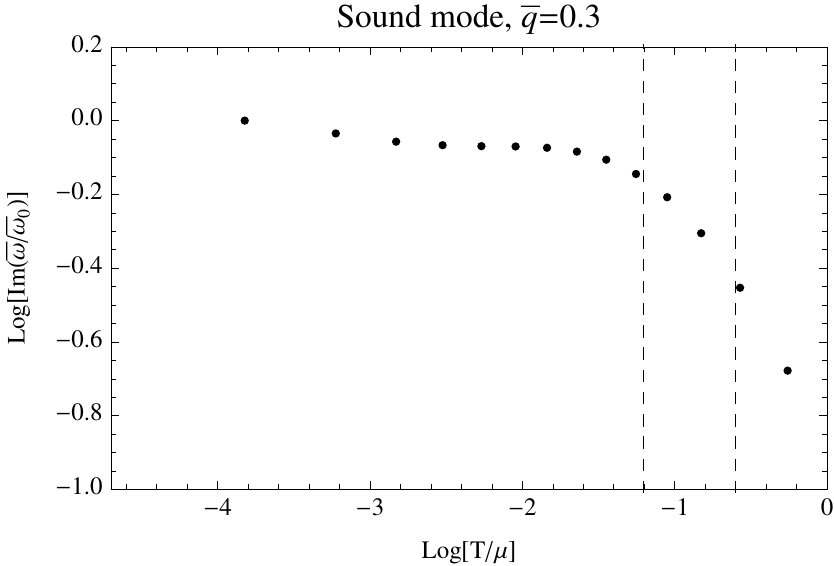}
\includegraphics[scale=0.88]{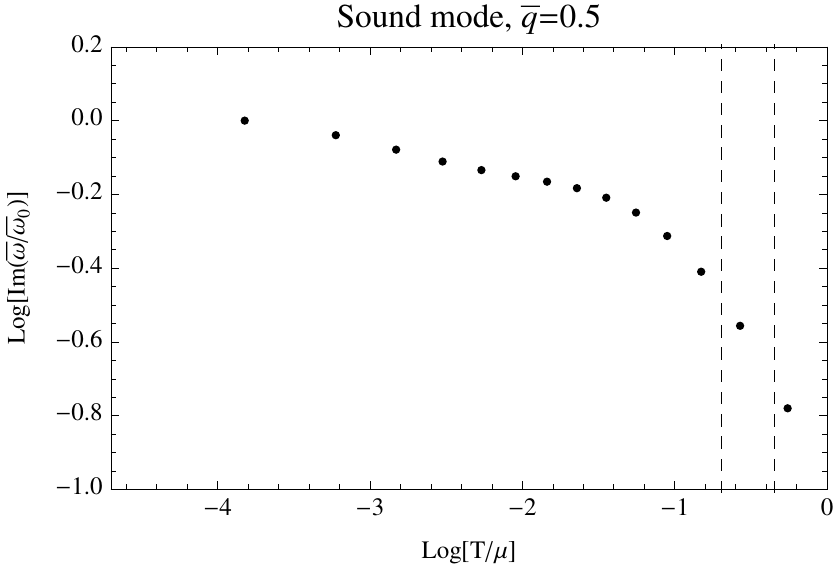}
\caption{Variation of the imaginary part of the sound mode as the temperature is increased, in the regime $T<\mu$. The dots are the numerical results for $T>0$, and the two dashed lines on each plot denote $T/\mu=q/\mu$ and $T/\mu=\sqrt{q/\mu}$ as one moves to the right along the plot.}
\label{fig:Tdependenceimaginarysoundlogarithmic}
\end{center}
\end{figure*}

\paragraph{}
To make an easier comparison with Landau Fermi liquid theory, we plot the temperature dependence of the imaginary part of the sound mode on a logarithmic scale in figure \ref{fig:Tdependenceimaginarysoundlogarithmic}. These plots show only the region $T<\mu$ where we may expect such a theory to apply, and the imaginary part of the sound mode is normalised by $\wbar_0$, its value at the lowest non-zero temperature we can access.

\paragraph{}
There is a stark contrast between these plots and the results expected for an LFL zero sound mode, shown in figure \ref{fig:LFLplot}.\footnote{The magnitude of the frequency of the sound mode in the RN-$AdS_4$ theory is of the same order as its momentum for the results shown, and thus we compare to the LFL results with $\omega\rightarrow q$.} Landau's theory predicts that as the temperature is increased at fixed $q$ and $\mu$, the imaginary part of the zero sound mode should be approximately constant up until $T/\mu\sim q/\mu$. Between $T/\mu\sim q/\mu$ and $T/\mu\sim\sqrt{q/\mu}$, it should increase like $T^2$. Above $T/\mu\sim\sqrt{q/\mu}$ and below $T/\mu\sim1$, it should decrease like $T^2$. None of these features are present in our results. The magnitude of the imaginary part of the sound mode in our theory shows no significant temperature dependence until $T\sim\mu$. Above this, it begins to approach the $\mu=0$, $\omega,q\ll T$ result (\ref{eq:herzogsound}) where it decreases as $1/T$. 

\begin{figure*}
\begin{center}
\includegraphics[scale=0.88]{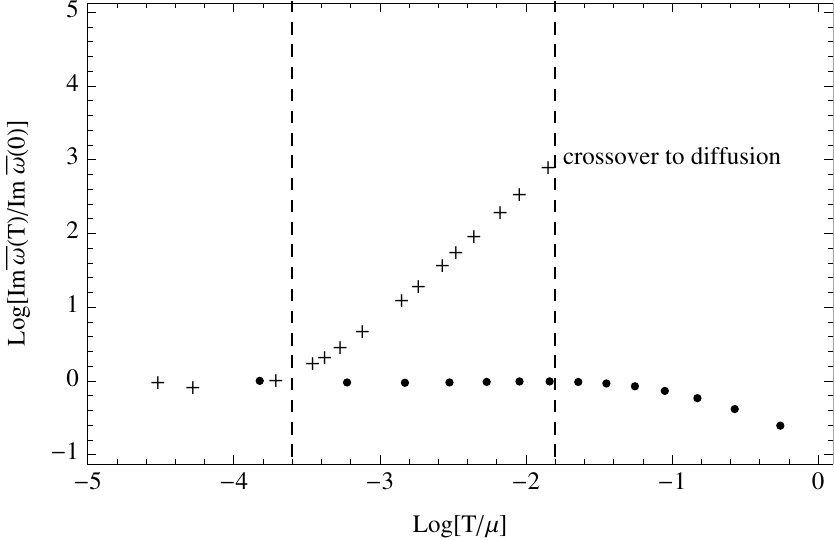}
\caption{A superposition of the plots of the temperature dependence of the normalised imaginary part of the sound mode when $q/\mu=0.2$ for both the D3/D7 theory and the RN-$AdS_4$ theory. Crosses denote the D3/D7 numerical results \protect\cite{Davison:2011ek} and circles denote the RN-$AdS_4$ results. Moving from left to right, the dotted lines mark the transition points between the quantum and thermal collisionless regimes, and the thermal collisionless regime and the hydrodynamic regime, in the D3/D7 theory. These occur when $\omega\sim T$ and $\omega\sim T^2/\mu$ respectively. There are no results for the D3/D7 sound mode in the hydrodynamic regime since the hydrodynamic sound mode is suppressed in the probe brane limit. We refer the reader to \protect\cite{Davison:2011ek} for a more detailed discussion of these features.}
\label{fig:soundcomparisonplot}
\end{center}
\end{figure*}

\paragraph{}An explicit comparison between these RN-$AdS_4$ results and the corresponding D3/D7 results is shown in figure \ref{fig:soundcomparisonplot}. This highlights the fact that the D3/D7 sound mode behaves like the LFL zero sound mode, whereas the RN-$AdS_4$ sound mode does not.
\section{Further temperature-dependent properties of the theory when $\qbar<1$}
\label{sec:furtherTdependentproperties}
\paragraph{}In addition to the sound modes at $T=0$,  there are other propagating modes lying deeper in the complex frequency plane as well as a branch cut along the negative imaginary frequency axis \cite{Edalati:2010pn}. In this section, we study how this configuration changes as $T/\mu$ is increased at a fixed momentum $\qbar<1$. In particular, we concentrate on the longest-lived purely imaginary mode - this exists at non-zero temperatures as the branch cut mentioned above dissolves into a series of poles when $T\ne0$. We note here that when $\mu=0$ and $\omega,q\ll T$, the longest-lived purely imaginary mode has the dispersion relation
\begin{equation}
\label{eq:herzogdiffusion}
\omega=-i\frac{3}{4\pi T}q^2+O(q^3),
\end{equation}
corresponding to hydrodynamic charge diffusion \cite{Herzog:2002fn}. We expect to recover this behaviour at non-zero $\mu$ in the limit $\mu\ll\omega,q\ll T$.

\paragraph{}
We also show in this section how the energy density and charge density spectral functions of the theory change with the temperature, and in particular how the residues of the long-lived modes play an important role in the transition from sound domination of the charge density spectral function to diffusion domination.

\subsection{Temperature dependence of the diffusion mode}
\label{sec:diffusionTdependence}
\begin{figure*}
\begin{center}
\includegraphics[scale=0.88]{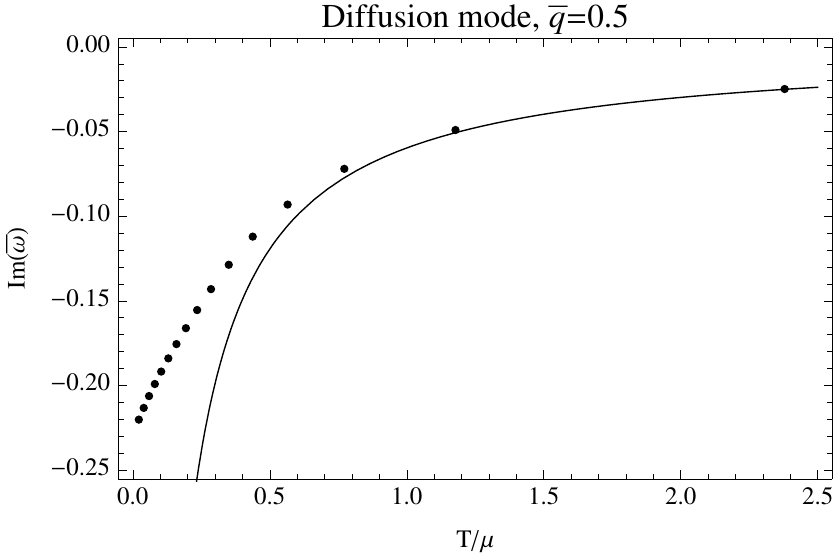}
\caption{Variation of the imaginary part of the longitudinal diffusion mode as the temperature is increased. The dots are the numerical results for $T>0$, and the solid line is the $\mu=0$ analytic result (\ref{eq:herzogdiffusion}).}
\label{fig:Tdependencediffusion}
\end{center}
\end{figure*}

\paragraph{}
We begin by studying the longitudinal diffusion mode of the theory. At $T=0$, the negative imaginary frequency axis is a branch cut \cite{Edalati:2010pn}. At non-zero temperatures, this branch cut dissolves into a series of poles along the axis and generically these become less stable (they recede into the complex plane) as the temperature is increased. However, one of the modes is special in that it becomes more stable as the temperature is increased, and at very high temperatures it becomes the $\mu=0$ hydrodynamic charge diffusion mode (\ref{eq:herzogdiffusion}). Figure \ref{fig:Tdependencediffusion} shows how the imaginary part of this mode changes with the temperature. Its real part is always zero. Unlike for the sound mode, this plot has the same shape for all values of $0.01\le\qbar\le0.5$ and so we show only one for brevity. The decay rate of this mode decreases monatonically as the temperature is increased, and is described well by the $\mu=0$ result (\ref{eq:herzogdiffusion}) when $T\gtrsim\mu$. Again, it is non-trivial that the $\mu=0$ result holds in the regime $q\ll\mu\ll T$. Note that there is no $T=0$ point on this plot because it does not make sense to ask where the pole is in that case - the whole negative imaginary frequency axis forms a branch cut.
\subsection{Movement of the poles in the complex frequency plane with temperature}
\begin{figure*}
\begin{center}
\includegraphics[scale=0.88]{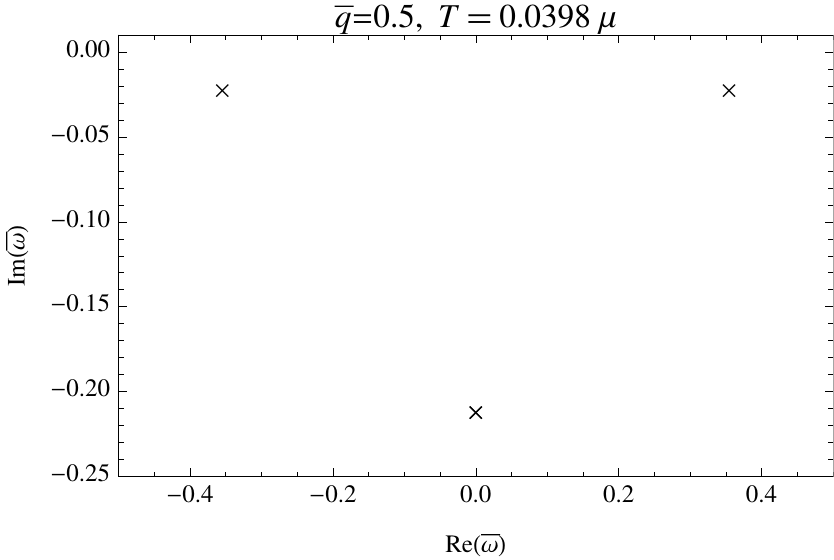}
\includegraphics[scale=0.88]{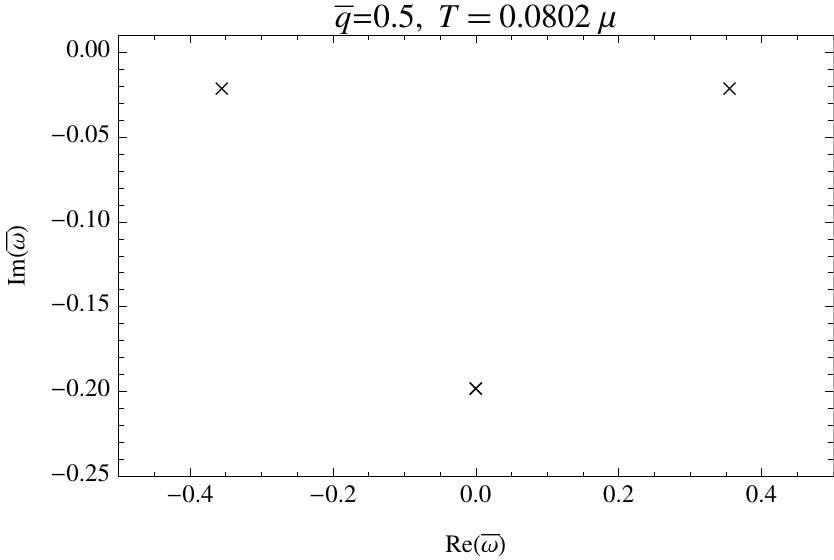}
\includegraphics[scale=0.88]{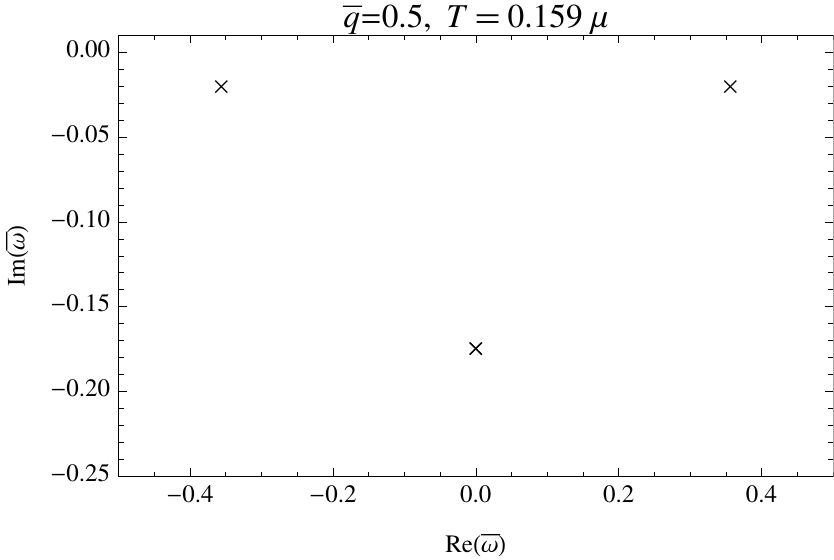}
\includegraphics[scale=0.88]{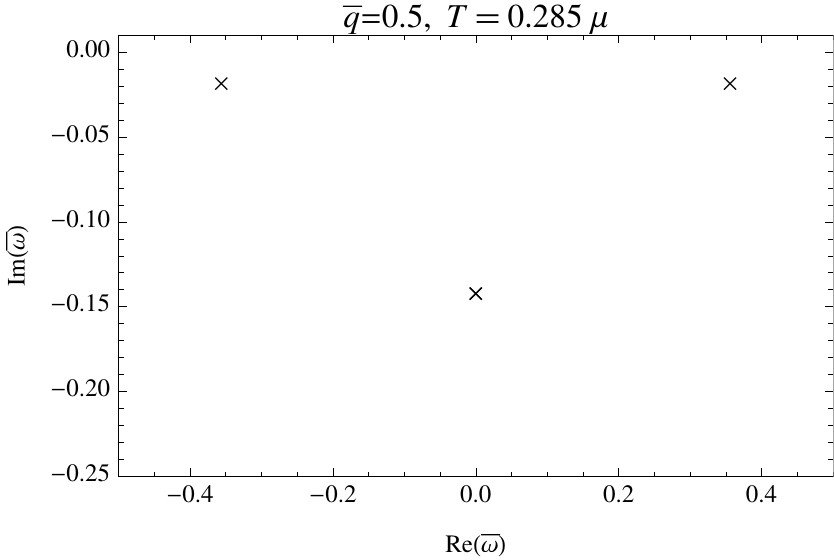}
\includegraphics[scale=0.88]{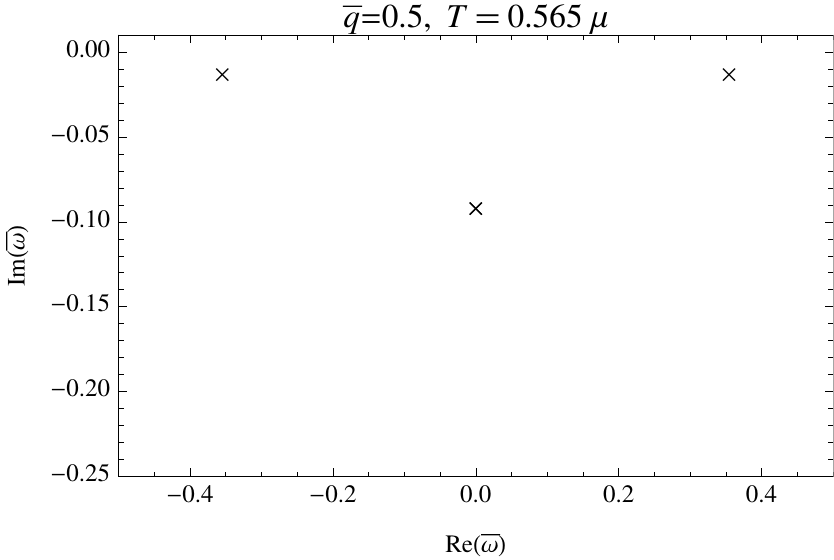}
\includegraphics[scale=0.88]{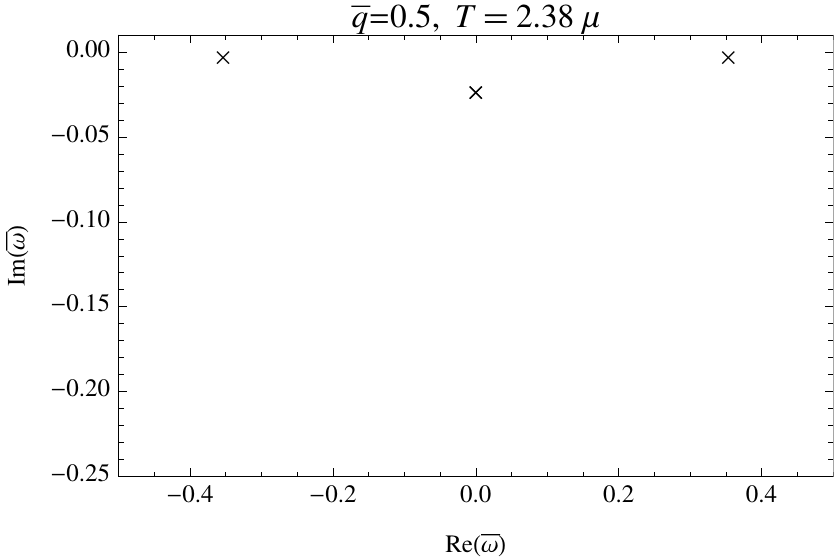}
\caption{Movement of the sound and diffusion poles in the complex frequency plane as the temperature is increased at fixed $\qbar=0.5$. An animated version of this figure is available at \url{http://www.physics.ox.ac.uk/users/Davison/RNAdS4animations1.html}.}
\label{fig:Tdependencecomplexplanehydro}
\end{center}
\end{figure*}

\paragraph{}
It is instructive to view the simultaneous movement of the Green's function poles described previously in the complex frequency plane as the temperature is increased - this is shown in figure \ref{fig:Tdependencecomplexplanehydro}.
As described previously, the sound and diffusion modes both become more stable, approaching the real axis as the temperature is increased. Note that the sound mode is closer to the real axis than the diffusion mode for all values of the temperature and thus it is always the longest-lived mode of the theory. However this does not mean that it always dominates the low-energy properties of the theory, as we shall show in the following subsection. Finally, we note that both the sound and diffusion poles coexist for all non-zero values of the temperature that we can access numerically ($T\gtrsim0.02\mu$). This is in contrast to the strongly-coupled D3/D7 field theory in which the low temperature sound poles collide to form the high temperature diffusion pole \cite{Davison:2011ek}.

\paragraph{}In addition to the sound and diffusion poles, there are `secondary' modes corresponding to poles lying deeper in the complex frequency plane. These are much shorter-lived than the sound and diffusion poles when $\qbar<1$ and become less stable as the temperature is increased. They do not have a significant effect on the low energy properties of the theory when $\qbar<1$ and hence we do not show their temperature dependence for brevity.
\subsection{Variation of the spectral functions with temperature}
\begin{figure*}
\begin{center}
\includegraphics[scale=0.88]{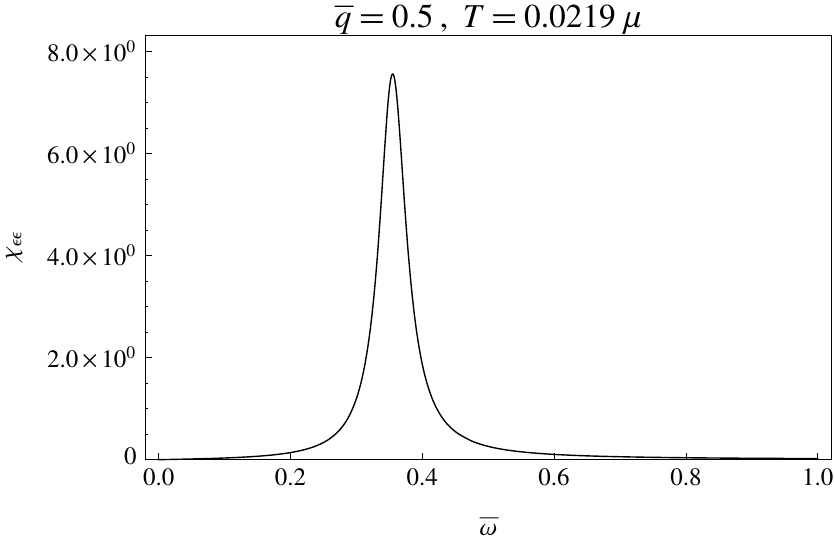}
\includegraphics[scale=0.88]{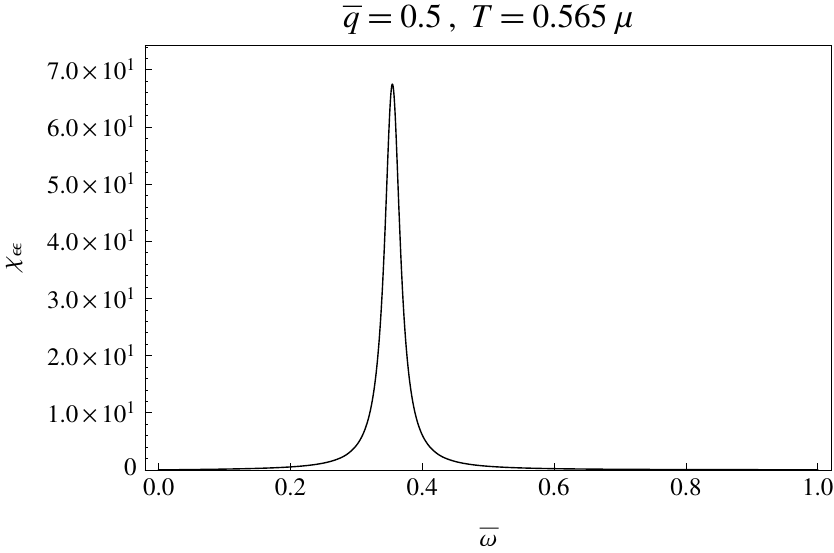}
\includegraphics[scale=0.88]{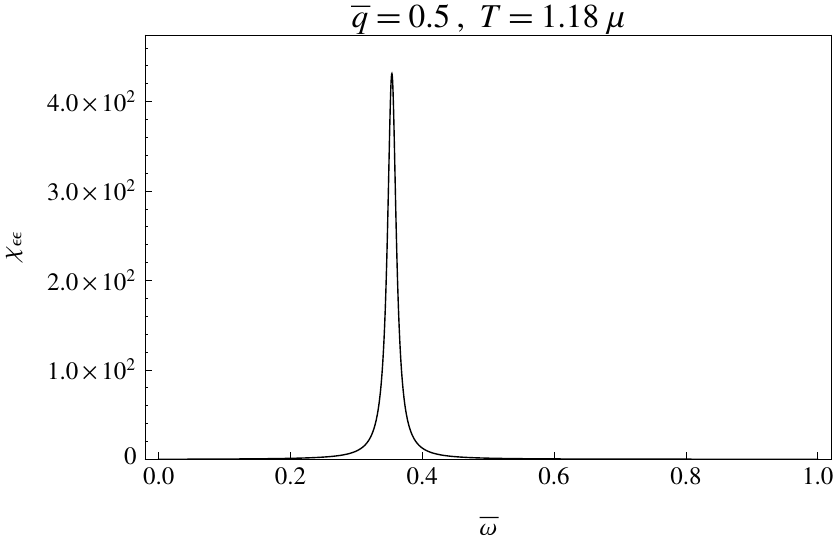}
\includegraphics[scale=0.88]{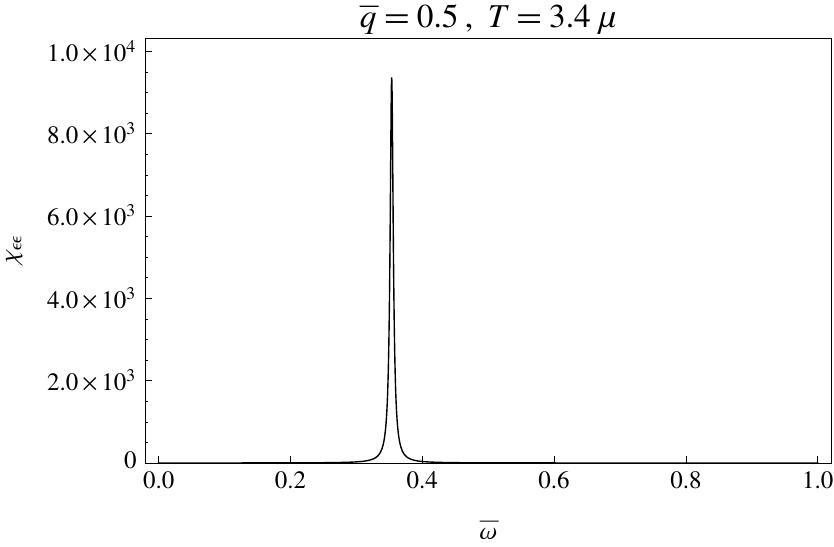}
\includegraphics[scale=0.88]{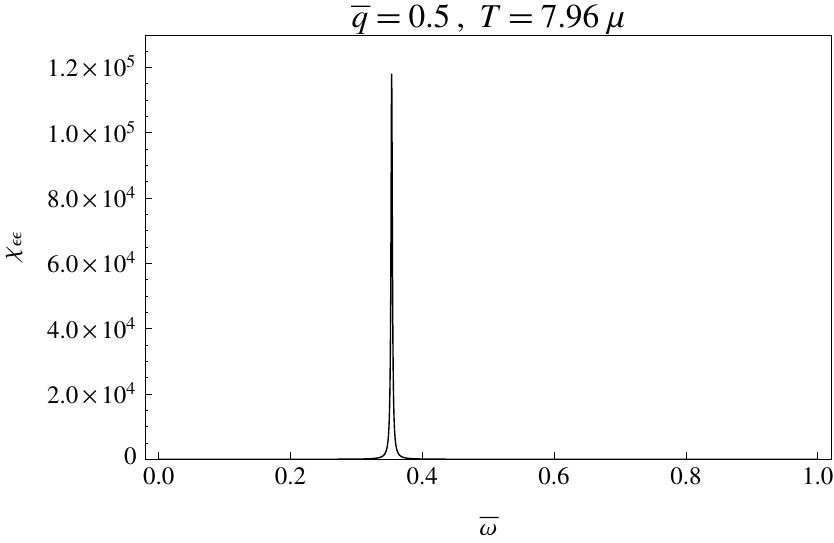}
\includegraphics[scale=0.88]{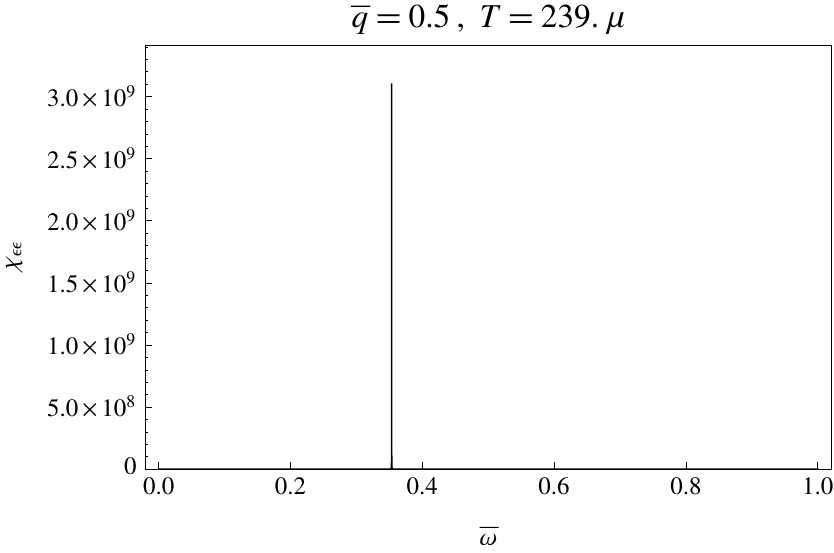}
\caption{The energy density spectral function for $\qbar=0.5$ as the temperature is increased, in units of $2\mu^2r_0/\kappa_4^2$. The peak due to sound propagation dominates at all temperatures. An animated version of this figure is available at \url{http://www.physics.ox.ac.uk/users/Davison/RNAdS4animations1.html}.}
\label{fig:Tdependencespectralfunctionsenergy}
\end{center}
\end{figure*}

\begin{figure*}
\begin{center}
\includegraphics[scale=0.88]{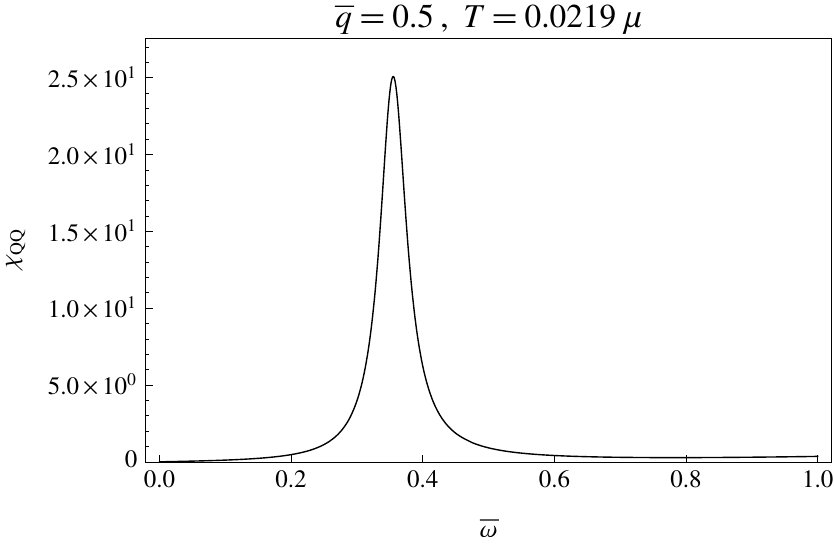}
\includegraphics[scale=0.88]{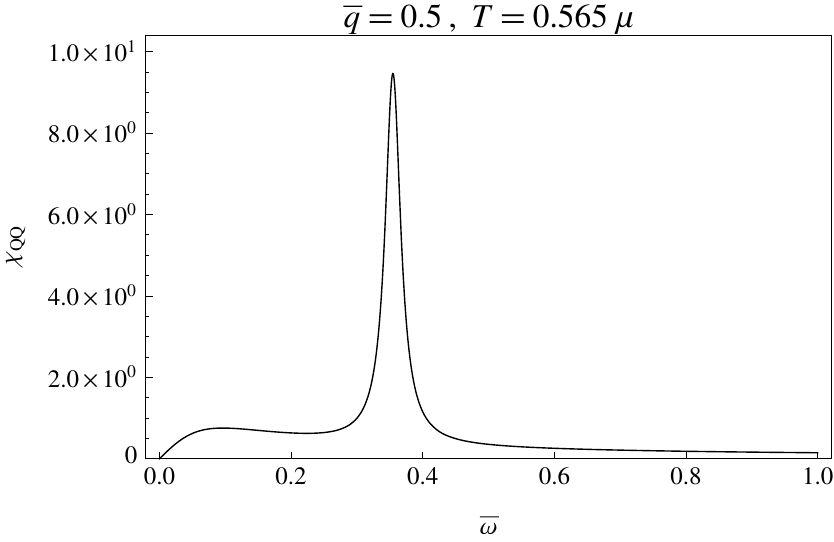}
\includegraphics[scale=0.88]{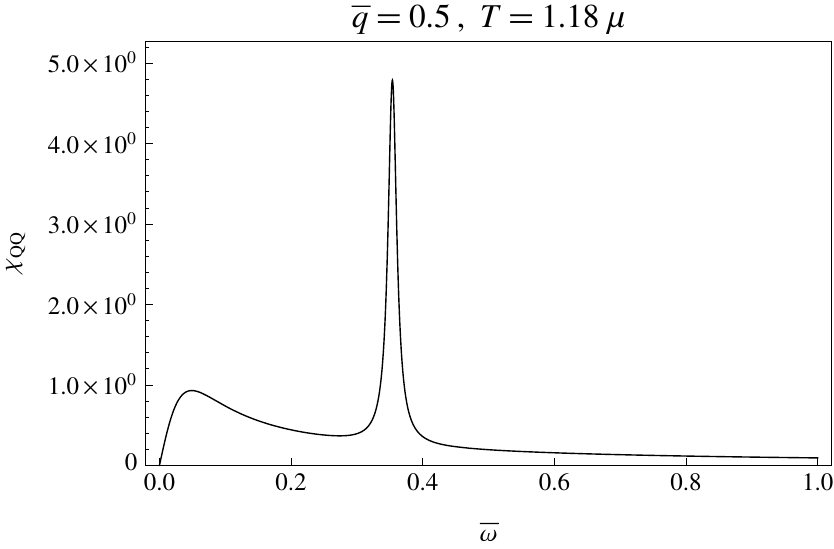}
\includegraphics[scale=0.88]{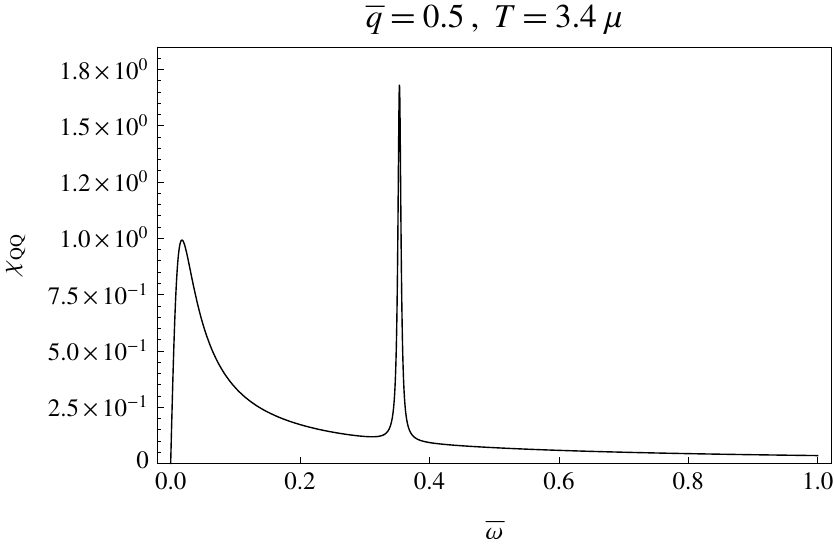}
\includegraphics[scale=0.88]{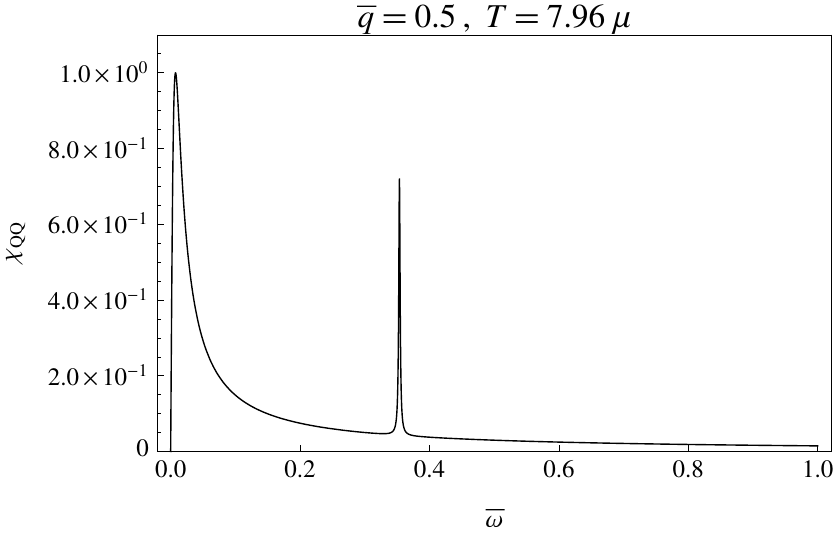}
\includegraphics[scale=0.88]{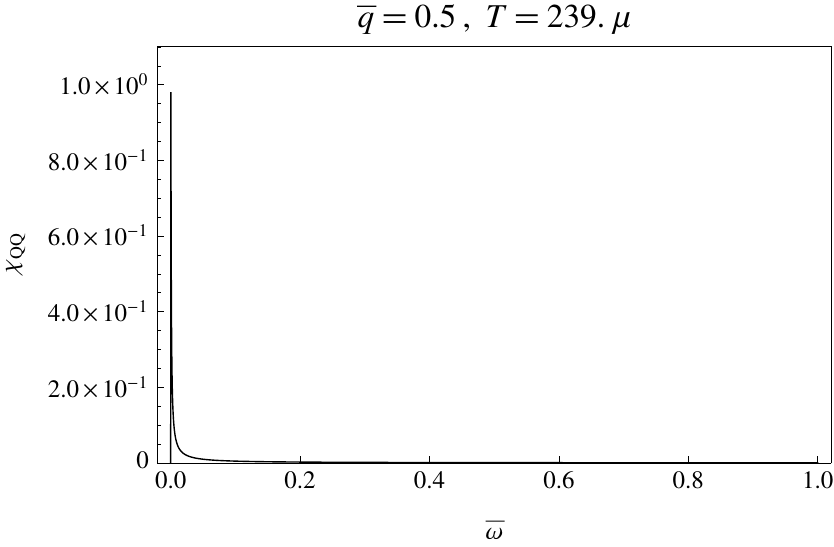}
\caption{The charge density spectral function for $\qbar=0.5$ as the temperature is increased, in units of $2r_0/\kappa_4^2$. There is a crossover between sound domination and diffusion domination at high temperature. An animated version of this figure is available at \url{http://www.physics.ox.ac.uk/users/Davison/RNAdS4animations1.html}.}\label{fig:Tdependencespectralfunctionscharge}
\end{center}
\end{figure*}

\begin{figure*}
\begin{center}
\includegraphics[scale=0.88]{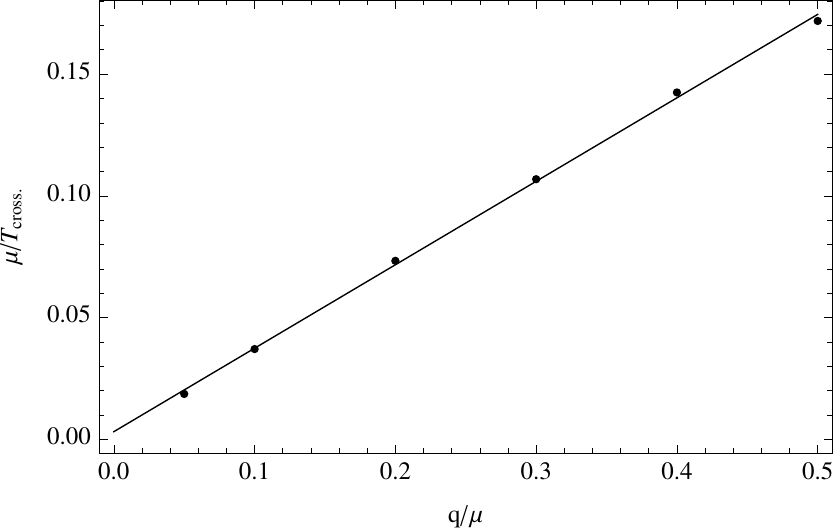}
\caption{The dependence of the crossover value of $\mu/T$ upon $q/\mu$. The best fit straight-line to these results has intercept $\approx0.003$ and gradient $\approx0.34$.}
\label{fig:spectralcrossoverdata}
\end{center}
\end{figure*}

\paragraph{}
Until now, we have focused exclusively on the positions of the poles of the retarded Green's functions in the complex frequency plane. While these are interesting, they do not tell us the full story of how charge and energy are transported in the theory. To investigate this, we now turn our attention to the spectral functions of the energy density and charge density. These determine the average work done on the system when an external source of some frequency is applied to either the energy density or the charge density respectively. Despite the fact that the retarded Green's functions of both of these operators have the same set of poles, their spectral functions are quite different as shown in figures \ref{fig:Tdependencespectralfunctionsenergy} and \ref{fig:Tdependencespectralfunctionscharge}. 
\paragraph{}
At very low temperatures, both spectral functions are dominated by the peak corresponding to sound propagation. As the temperature is increased, the spectral function of the energy density undergoes a fairly unremarkable change - the sound peak becomes narrower and taller (corresponding to a longer-lived excitation) but completely dominates at all temperatures. In contrast to this, the sound peak of the charge density spectral function becomes smaller (and narrower) as the temperature increases. At a sufficiently high temperature it becomes so small that the peak around $\wbar=0$, corresponding to the high temperature diffusion mode, dominates the spectral function. At what temperature does this crossover occur? In figure \ref{fig:spectralcrossoverdata}, we show how the crossover value of $\mu/T$ (i.e. the value where the sound and diffusion peaks are of the same height) varies with $q/\mu$.
There is a clear linear relationship, signifying that the crossover occurs when
\begin{equation}
T_{\text{cross.}}\sim\mu^2/q,
\end{equation}
and that diffusion dominates when $T\gg\mu^2/q$. Note that since we are studying the range $\qbar<1$, this condition automatically implies that $T\gg\mu$. This crossover is reminiscent of the $\mu\rightarrow0$ limit. In that limit, the fluctuations of $T^{\mu\nu}$ and $J^\mu$ decouple resulting in a hydrodynamic sound pole in the $T^{\mu\nu}$ correlators and a hydrodynamic diffusion pole in the $J^\mu$ correlators. The charge density spectral function in this limit is shown in \cite{Herzog:2007ij}. However, we cannot interpret the crossover shown above to be due to approaching this limit, since it corresponds to the limit $\mu\ll q,T$ of our results, whereas the regime we are studying here is $q\ll\mu\ll T$.

\paragraph{}
In the D3/D7 theory, the corresponding crossover occurred for $T_{\text{cross.}}\sim\sqrt{q\mu}$ and was reminiscent of the collisionless/hydrodynamic crossover in an LFL \cite{Davison:2011ek}, but we do not have a similar explanation here. In particular, the crossover observed here in the RN-$AdS_4$ theory occurs outside of the `quantum liquid' regime $T\ll\mu$.
\section{Dispersion relations at fixed temperature $T<\mu$}
\label{sec:qdependenceresults}
\paragraph{}
In the previous sections, we established how an increase in temperature affects the sound and diffusion modes that exist at some fixed, low momentum $q\ll\mu$. We now turn our attention to studying the dispersion relations of these modes at a fixed temperature $T$ and chemical potential $\mu$. This is done by fixing $T/\mu$ and varying $\qbar$.
\subsection{The sound mode dispersion relation}
\paragraph{}
In \cite{Edalati:2010pn}, the dispersion relation at $T=0$ and $\qbar\ll1$ was found numerically to be of the form (\ref{eq:sounddispersion}) with $v_s=1/\sqrt{2}$ and $\Gamma_0=0.083$, which is remarkably close to the dispersion relation expected from the `zero temperature hydrodynamics' described in the introduction and in \cite{Edalati:2010pn}, which has $\Gamma_0=0.072$ rather than $0.083$.

\begin{figure*}
\begin{center}
\includegraphics[scale=0.9]{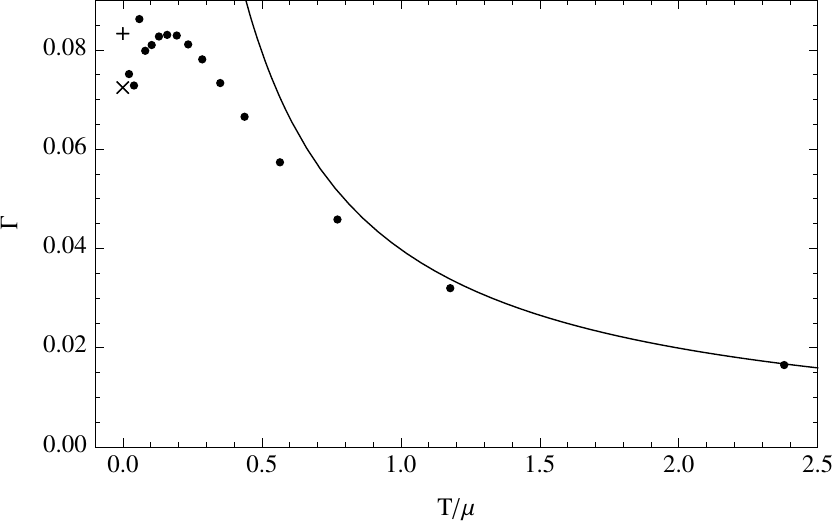}
\caption{The temperature dependence of the quadratic term $\Gamma$ in the imaginary part of the sound dispersion relation (\ref{eq:sounddispersionrelationallT}). Circles show our numerical results, the solid line shows the $\mu=0$ analytic result (\ref{eq:herzogsound}), the `$+$' shows the $T=0$ numerical result of \protect\cite{Edalati:2010pn} and the `$\times$' shows the prediction of `$T=0$ hydrodynamics'.}
\label{fig:sounddispersionfit}
\end{center}
\end{figure*}

\begin{figure*}
\begin{center}
\includegraphics[scale=0.88]{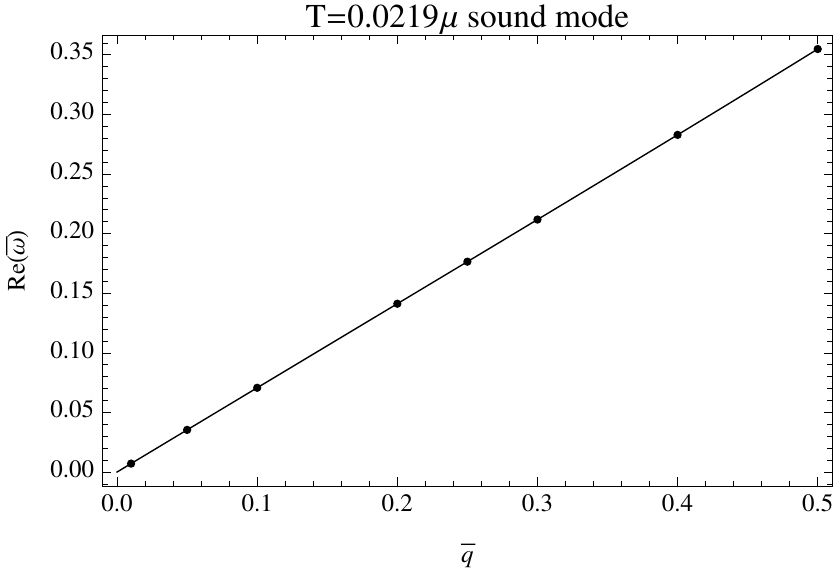}
\includegraphics[scale=0.88]{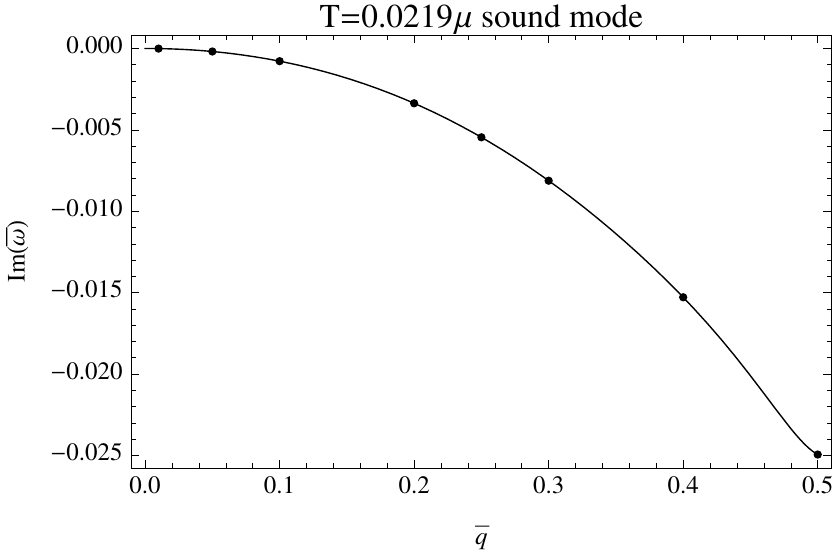}
\caption{The dispersion relation of the sound mode at $T=0.0219\mu$ for $0.01\leq\qbar\leq0.5$. The circles show our numerical results and the solid line is the best fit $\wbar\approx\qbar/\sqrt{2}-i0.075\qbar^2+O(\qbar^3)$.}
\label{fig:sounddispersionfitlowT}
\end{center}
\end{figure*}

\paragraph{}We found a dispersion relation of the form (\ref{eq:sounddispersionrelationallT}) to be valid for non-zero temperatures also. The quadratic coefficient of the attenuation $\Gamma$ as a function of temperature is shown in figure \ref{fig:sounddispersionfit}. These results were obtained by fitting over the range $0.01\le\qbar\le0.5$, and an example of the fit for $T=0.0219\mu$ is shown in figure \ref{fig:sounddispersionfitlowT}. 
At high temperatures $T\gg\mu$, it agrees with the $\mu=0$ result (\ref{eq:herzogsound}), and at low temperatures $T\ll\mu$ it approaches a similar value to the $T=0$ result of \cite{Edalati:2010pn}. Although our results at very low $T$ do not match smoothly onto the $T=0$ numerical result of \cite{Edalati:2010pn}, they differ only by around $10\%$ and we believe that this is most likely caused by numerical inaccuracies, which grow as the temperature is lowered. The proximity of the numerical results to the prediction of `$T=0$ hydrodynamics' is surprising - ultimately, an analytic calculation will be needed to determine whether these small discrepancies are due to inaccurate numerics, or whether this proximity is in fact a coincidence. The general trend of the results is clear however - as the temperature increases, the sound mode becomes more stable, as was observed in section \ref{sec:Tdependencesound}.

\begin{figure*}
\begin{center}
\includegraphics[scale=0.88]{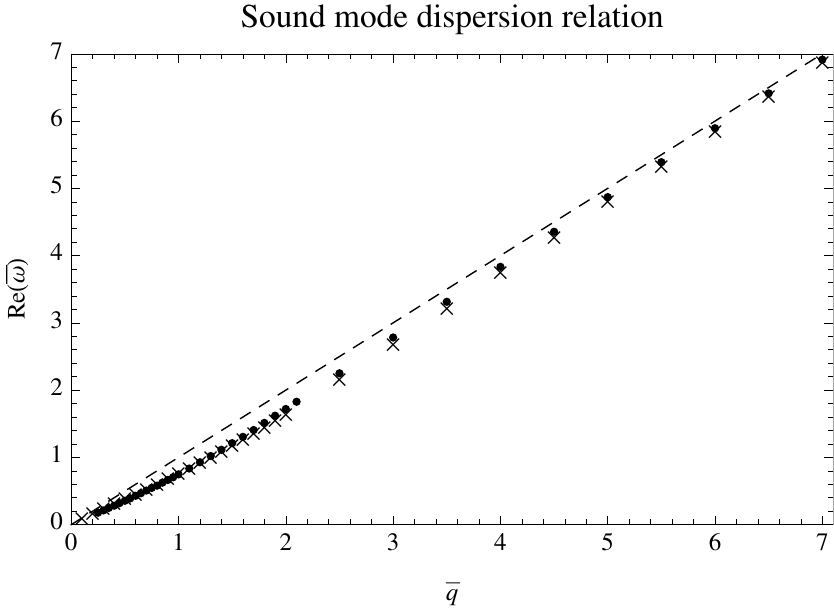}
\includegraphics[scale=0.88]{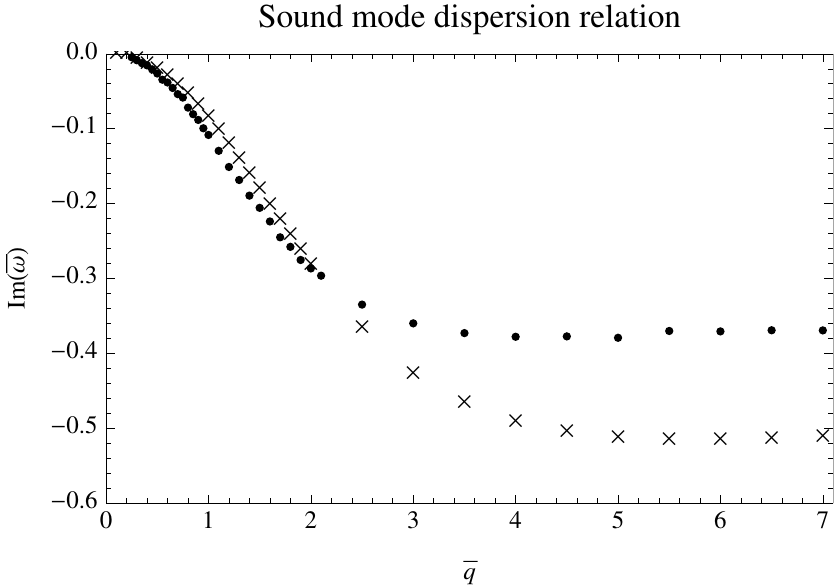}
\caption{The dispersion relation of the sound mode at two different temperatures: $T=0$ (circles) and $T=0.159\mu$ (crosses). The dashed line is the line $\text{Re}\left(\wbar\right)=\qbar$.}
\label{fig:qdependencesound}
\end{center}
\end{figure*}

\paragraph{}
When $\qbar\gtrsim1$, this series form of the dispersion relation is useless. Figure \ref{fig:qdependencesound} shows the dispersion relation of the sound mode when $q>\mu>T$, at two different temperatures: $T=0$ and $T=0.159\mu$.
The dispersion relations have the same shape at both temperatures - the real part asymptotes to $\wbar=\qbar$, and the imaginary part tends to a constant, in the region $q\gg\mu>T$.
\subsection{The diffusion mode dispersion relation}
\begin{figure*}
\begin{center}
\subfloat[Numerical results]{\label{fig:qdependencediffusion}\includegraphics[scale=0.88]{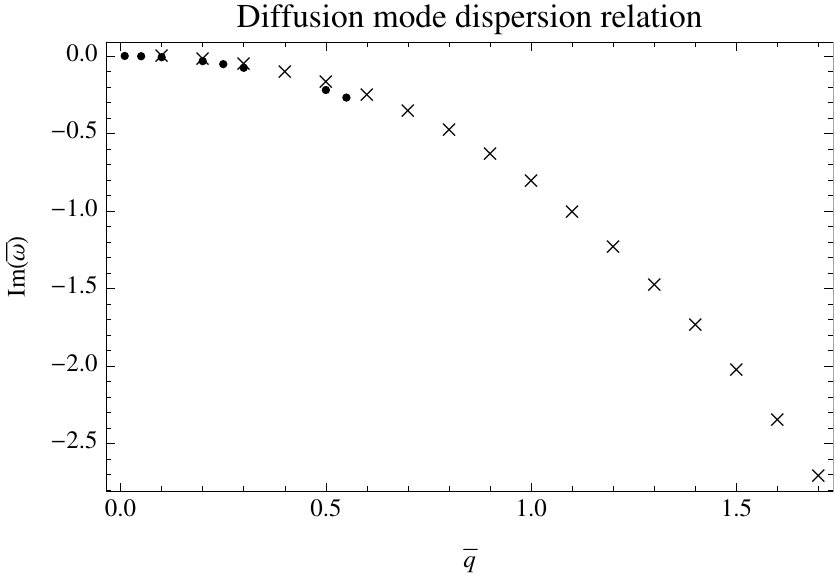}}
\subfloat[Best fit]{\label{fig:qdependencediffusionbestfit}\includegraphics[scale=0.88]{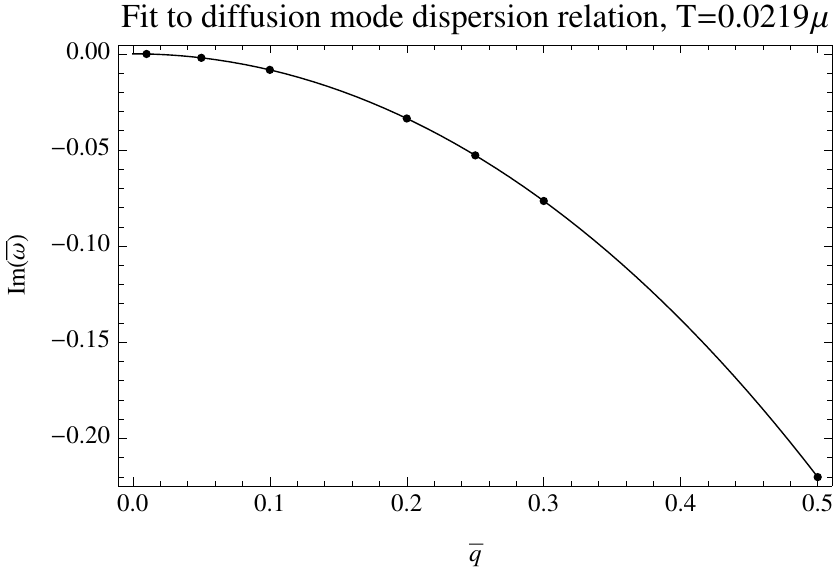}}
\caption{The dispersion relation of the diffusion mode at two different temperatures: $T=0.0219\mu$ (circles) and $T=0.159\mu$ (crosses) with the polynomial best fit at $T=0.0219\mu$ shown also (solid line). We cannot track the $T=0.0219\mu$ mode for as high momenta as the $T=0.159\mu$ mode.}
\end{center}
\end{figure*}

\paragraph{}
Recall that at non-zero temperatures, the branch cut along the negative imaginary frequency axis becomes a series of poles and that the most stable of these becomes the $\mu=0$ diffusion mode at high temperatures. Figure \ref{fig:qdependencediffusion} shows the imaginary part of the dispersion relation of this pole at two fixed, low temperatures $T=0.0219\mu$ and $T=0.159\mu$ (its real part is always zero).
At both temperatures, the pole recedes quickly into the complex plane as the momentum is increased. Performing a polynomial fit to the imaginary part in the range $0.01\leq\qbar\leq0.5$ at the very low temperature $T=0.0219\mu$, we found a dispersion relation of the form (\ref{eq:diffusiondispersionrelationallT}) with $D\approx0.83$. The fit is shown in figure \ref{fig:qdependencediffusionbestfit}. This therefore is an analogue, at low temperatures $T<q<\mu$, of the $\mu=0$, $q\ll T$ hydrodynamic charge diffusion mode. 

\begin{figure*}
\begin{center}
\includegraphics[scale=0.9]{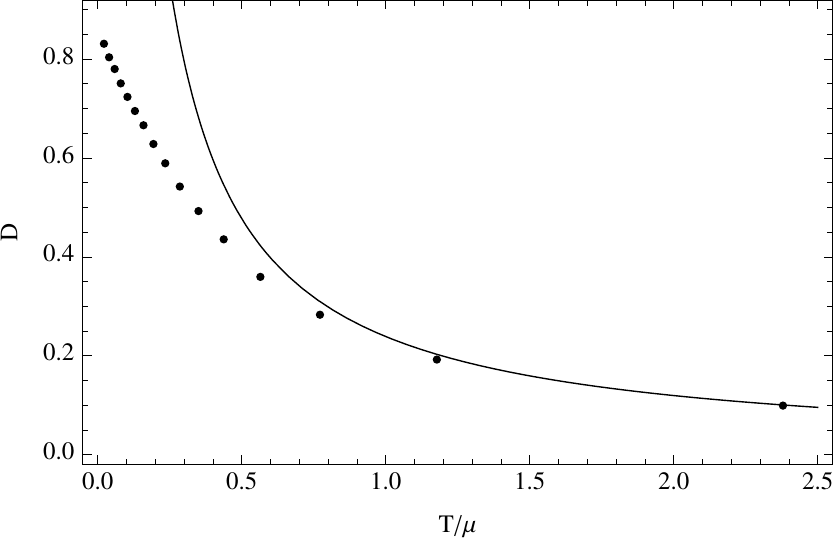}
\caption{The temperature dependence of the quadratic coefficient $D$ in the dispersion relation of the diffusion mode (\ref{eq:diffusiondispersionrelationallT}). The circles show our numerical results and the solid line is the analytic $\mu=0$ result (\ref{eq:herzogdiffusion}).}
\label{fig:TdependenceofD}
\end{center}
\end{figure*}

\paragraph{}In fact, this quadratic form of the dispersion relation (\ref{eq:diffusiondispersionrelationallT}) is valid for all non-zero temperatures that we could access - the dependence of $D$ upon $T$ is shown in figure \ref{fig:TdependenceofD}. We extracted $D$ from a fit over the range $0.01\le\qbar\le0.5$. It decreases monatonically as the temperature is raised, in agreement with the results of section \ref{sec:diffusionTdependence}, and approaches the $\mu=0$ result (\ref{eq:herzogdiffusion}) in the limit $T\gg\mu$. Again, we note that this is despite the fact that we are studying the regime $\mu\gg q$.
\paragraph{}
We could not obtain the numerical accuracy required to access non-zero temperatures lower than $T=0.0219\mu$ and hence we cannot say whether the mode exists with a quadratic dispersion relation for arbitrarily low non-zero temperatures. We emphasise again that this mode does not exist at $T=0$ itself (unlike the $T=0$ `R-spin diffusion' mode of \cite{Ammon:2011hz}), as there is a branch cut in that case.
\subsection{Dispersion relations of the secondary modes}
\begin{figure*}
\begin{center}
\includegraphics[scale=0.88]{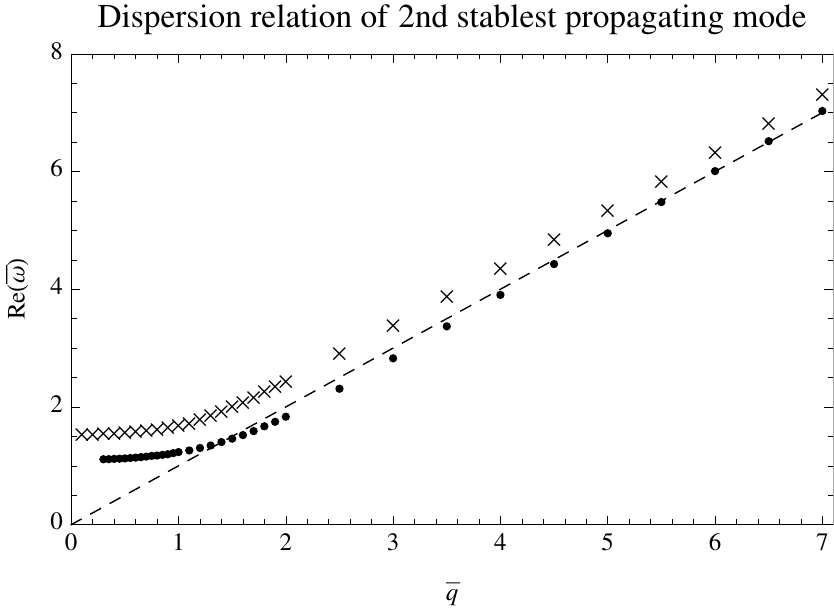}
\includegraphics[scale=0.88]{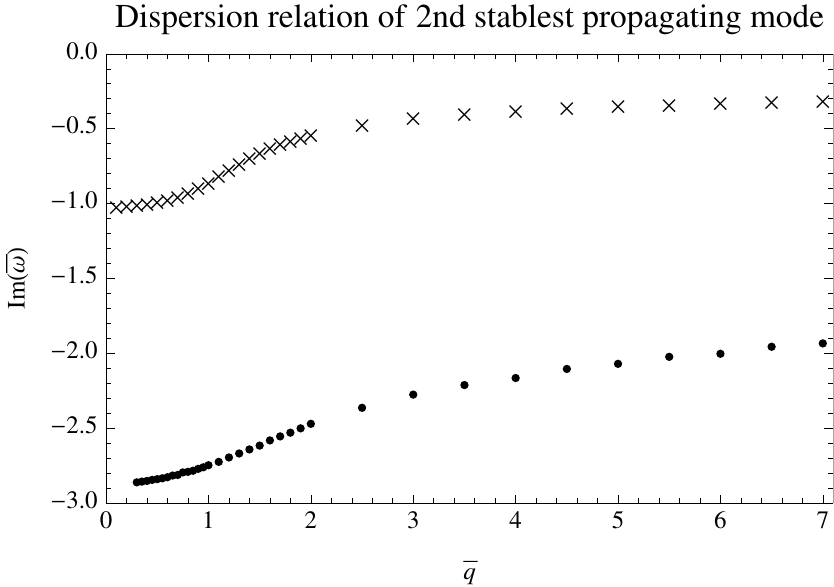}
\caption{The dispersion relation of the second stablest propagating mode at two different temperatures: $T=0$ (circles) and $T=0.159\mu$ (crosses). The dashed line is the line $\text{Re}\left(\wbar\right)=\qbar$.}
\label{fig:qdependencesecondarypropagating}
\end{center}
\end{figure*}

\begin{figure*}
\begin{center}
\includegraphics[scale=0.88]{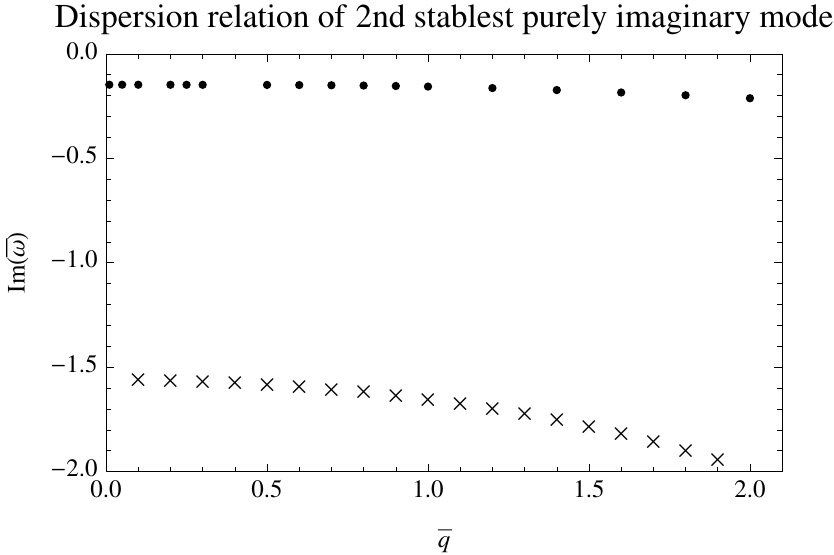}
\caption{The dispersion relation of the second-stablest purely imaginary mode at two different temperatures: $T=0.0219\mu$ (circles) and $T=0.159\mu$ (crosses).}
\label{fig:qdependencesecondaryimaginary}
\end{center}
\end{figure*}

\paragraph{}
Figures \ref{fig:qdependencesecondarypropagating} and \ref{fig:qdependencesecondaryimaginary} show the dispersion relations of the second stablest (`secondary') propagating and purely imaginary modes at two different temperatures. 
These differ qualitatively from the sound and diffusion modes described above in that $\wbar\ne0$ when $\qbar=0$. At high momenta $q\gg\mu>T$, the propagating secondary modes have the same form as the sound modes - their real parts asymptote to $\text{Re}\left(\wbar\right)=\qbar$ and their imaginary parts tend to a constant. The purely imaginary secondary mode recedes into the complex plane as the momentum is increased, and the rate at which this happens increases with the temperature.
\subsection{Movement of the poles in the complex frequency plane with momentum}
\begin{figure*}
\begin{center}
\includegraphics[scale=0.88]{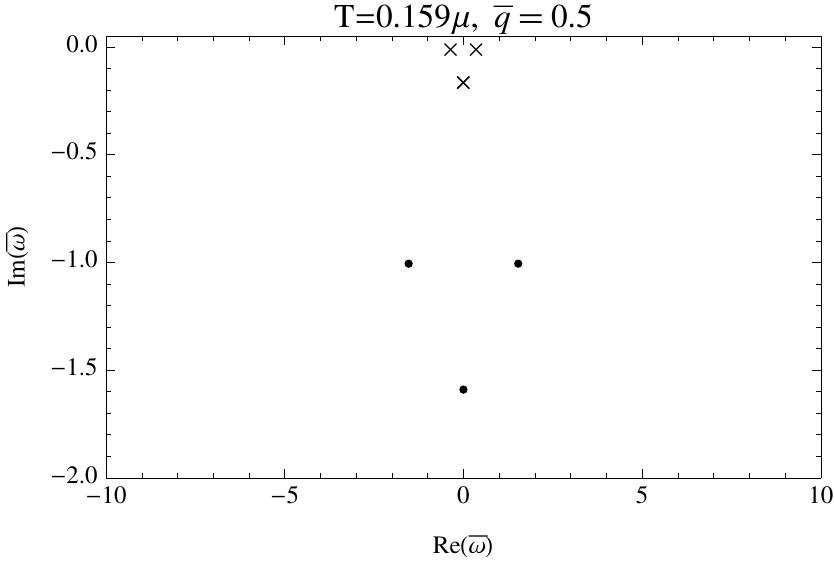}
\includegraphics[scale=0.88]{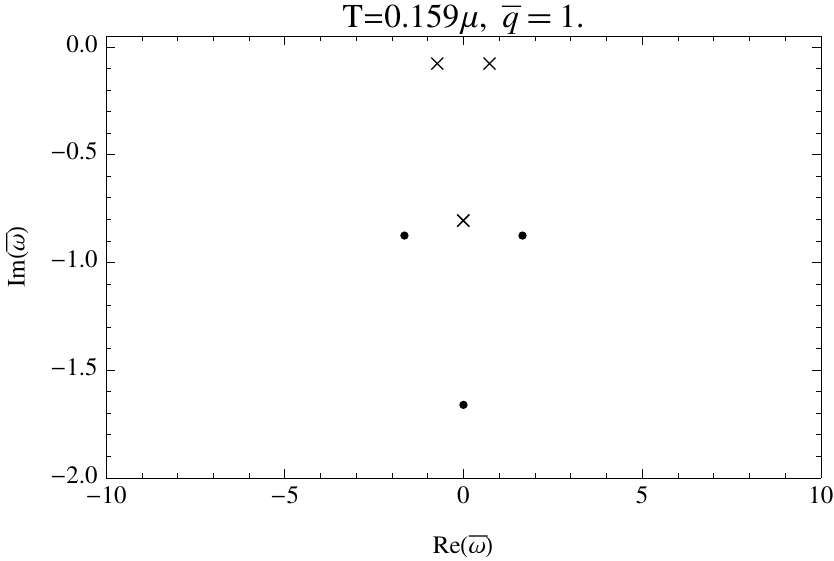}
\includegraphics[scale=0.88]{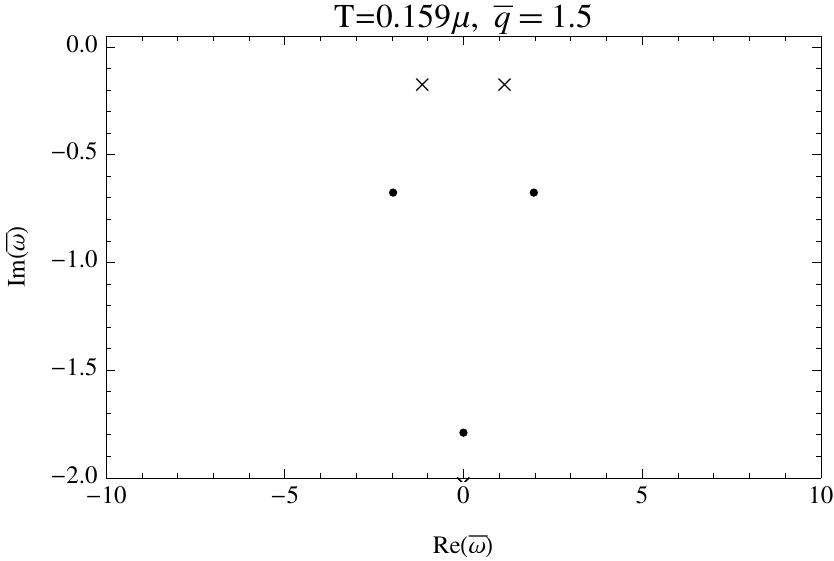}
\includegraphics[scale=0.88]{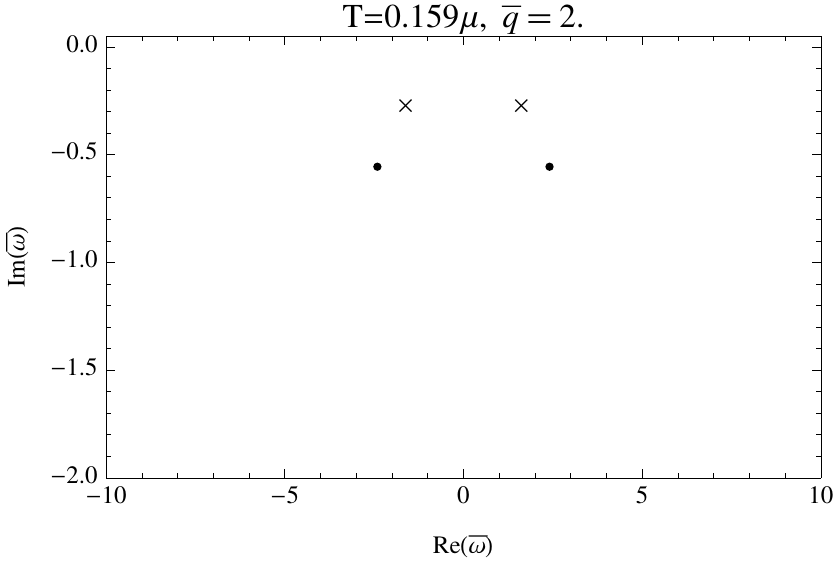}
\includegraphics[scale=0.88]{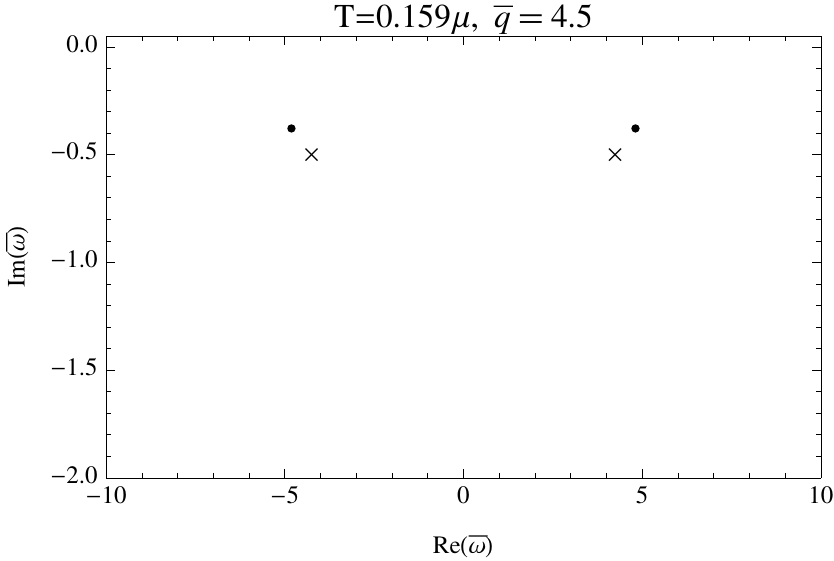}
\includegraphics[scale=0.88]{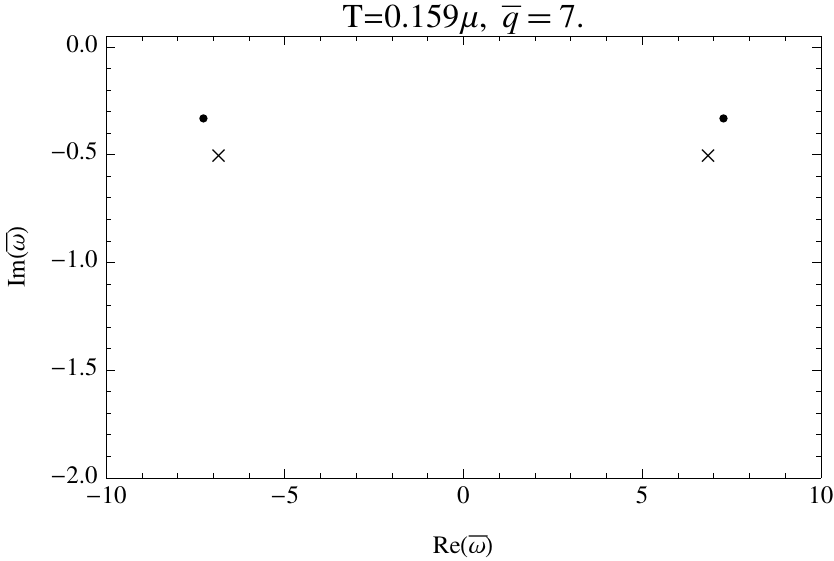}
\caption{Movement of the six longest-lived modes in the complex frequency plane as a function of momentum, for fixed $T=0.159\mu$. The crosses denote the sound and diffusion modes, and the circles denote the secondary propagating and imaginary modes. An animated version of this figure is available at \url{http://www.physics.ox.ac.uk/users/Davison/RNAdS4animations2.html}.}
\label{fig:qdependencecomplexplane}
\end{center}
\end{figure*}

\paragraph{}
It is instructive to view the simultaneous movements of these poles in the complex frequency plane as the momentum is increased. This is shown in figure \ref{fig:qdependencecomplexplane} for $T=0.159\mu$.
We see that as $\qbar$ is increased, the purely imaginary modes both become less stable as described previously. This figure shows that the diffusion mode destabilises much quicker than the secondary imaginary mode. The propagating modes show a different behaviour - their speeds both increase but their imaginary parts move in opposite directions and begin to approach each other in the complex plane as $\qbar$ increases. They eventually cross, before moving off horizontally together along the relativistic trajectory $\text{Re}\left(\wbar\right)=\qbar$. At these high values of $\qbar$ it is clear that our original separation of modes into the stablest (sound and diffusion), second stablest etc. is of no value.
\subsection{Variation of the spectral functions with momentum}
\begin{figure*}
\begin{center}
\includegraphics[scale=0.88]{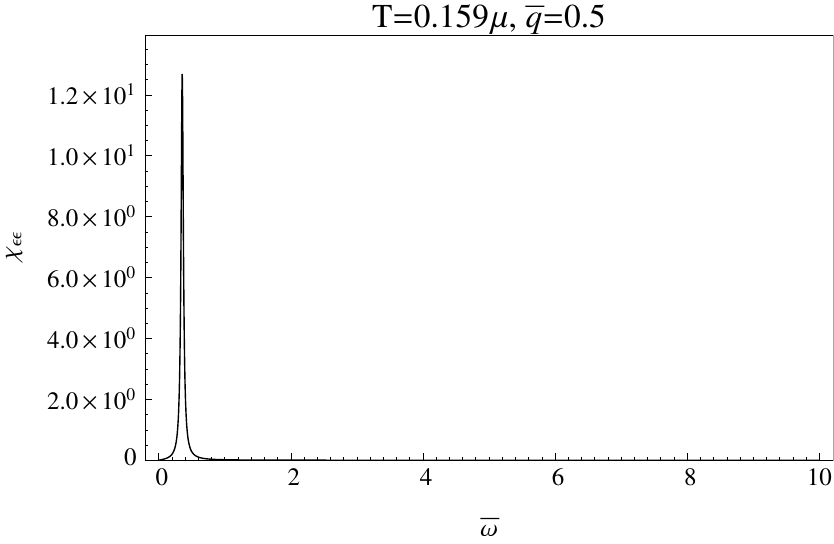}
\includegraphics[scale=0.88]{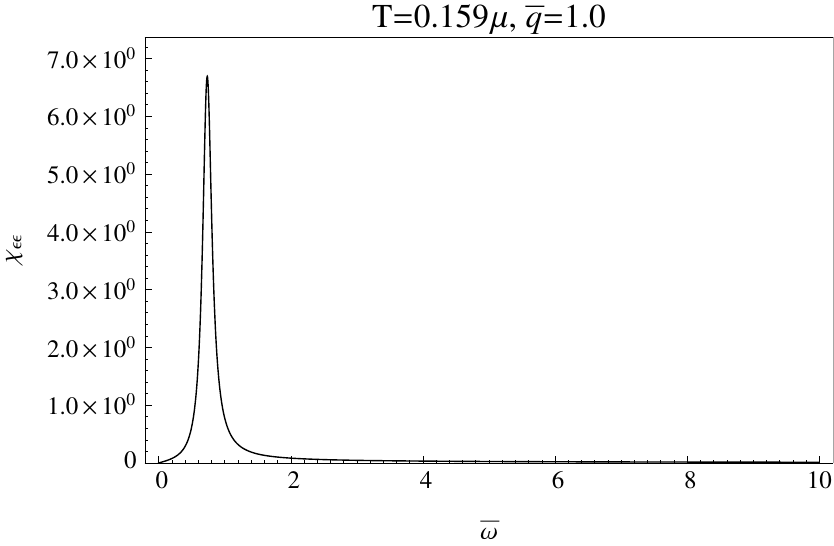}
\includegraphics[scale=0.88]{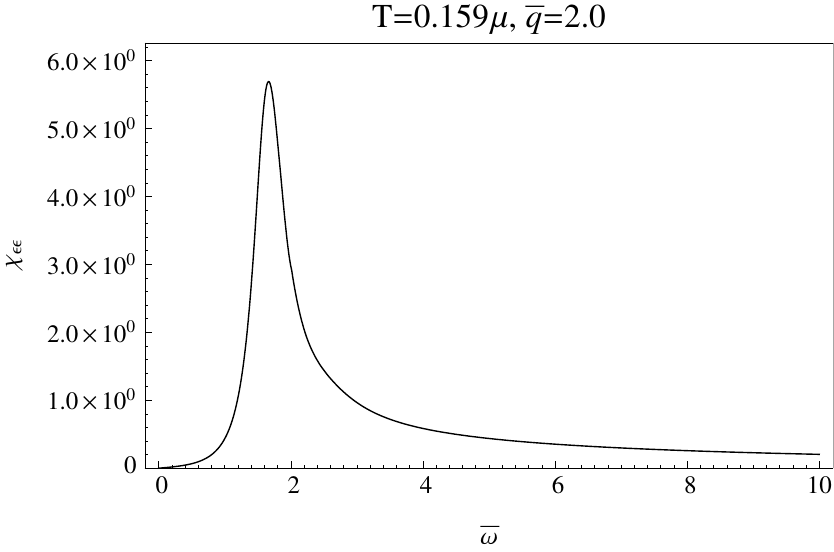}
\includegraphics[scale=0.88]{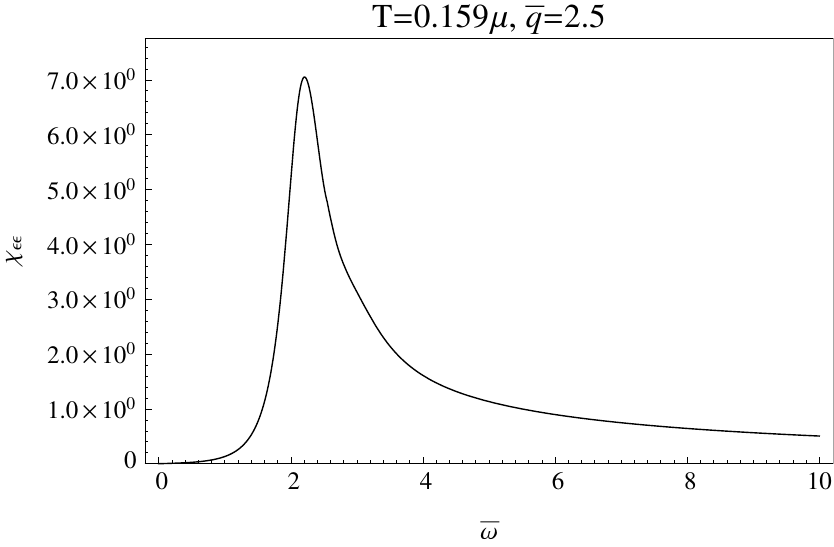}
\includegraphics[scale=0.88]{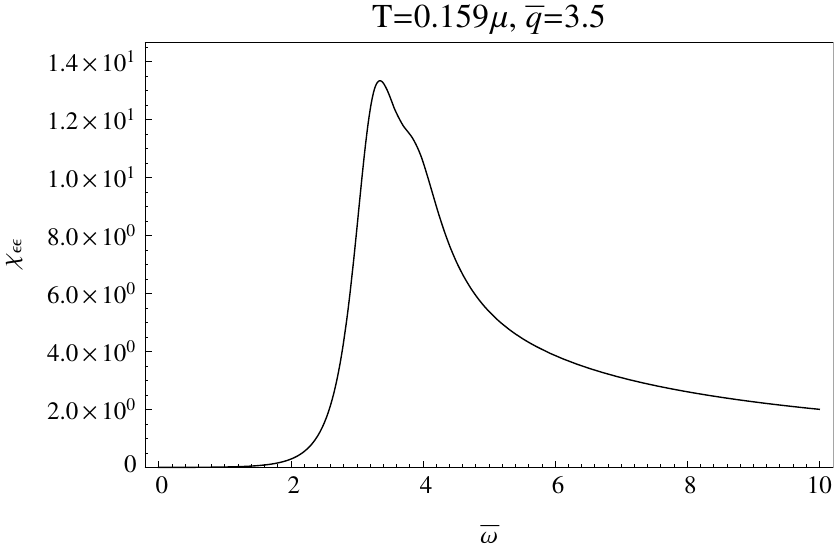}
\includegraphics[scale=0.88]{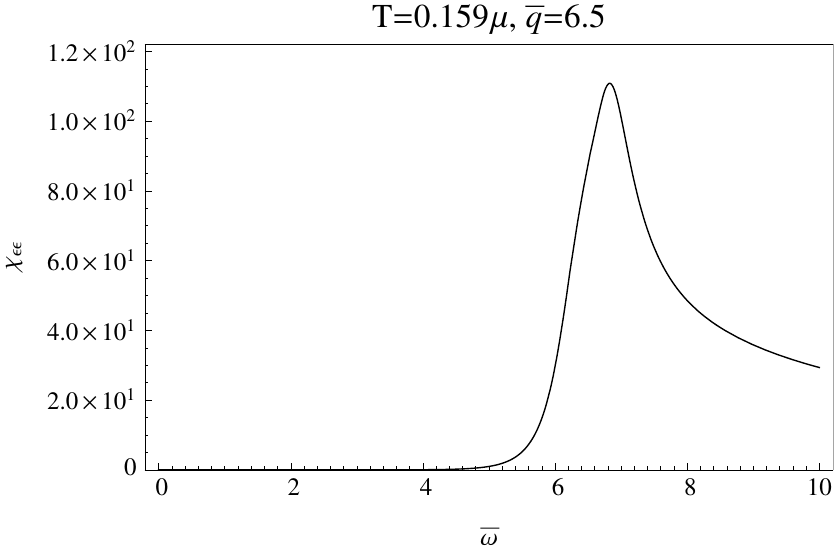}
\caption{The energy density spectral function for $T=0.159\mu$ as the momentum is increased, in units of $2\mu^2r_0/\kappa_4^2$. As the momentum is increased, the peak due to sound propagation becomes less dominant. An animated version of this figure is available at \url{http://www.physics.ox.ac.uk/users/Davison/RNAdS4animations2.html}.}
\label{fig:qdependencespectralfunctionsenergy}
\end{center}
\end{figure*}

\begin{figure*}
\begin{center}
\includegraphics[scale=0.88]{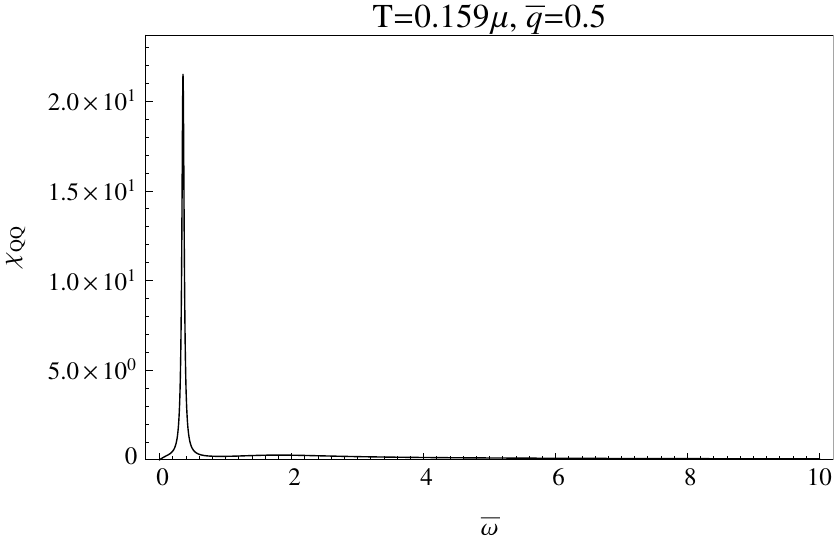}
\includegraphics[scale=0.88]{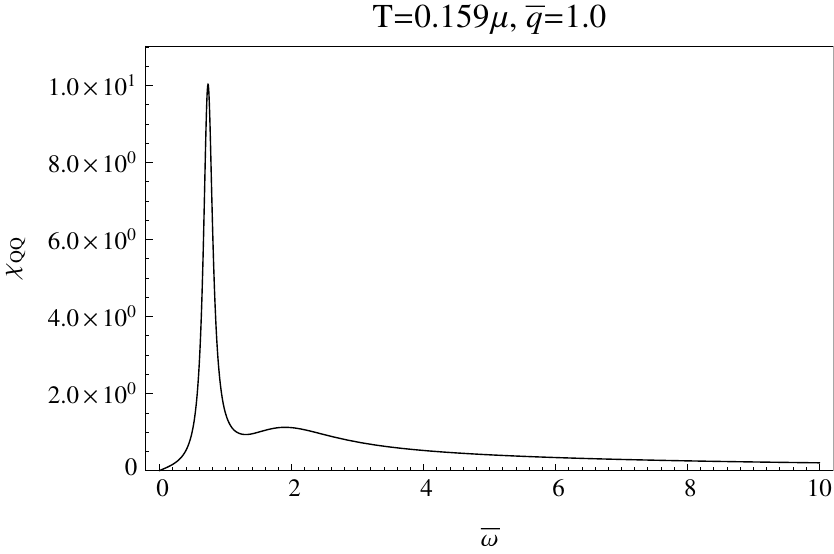}
\includegraphics[scale=0.88]{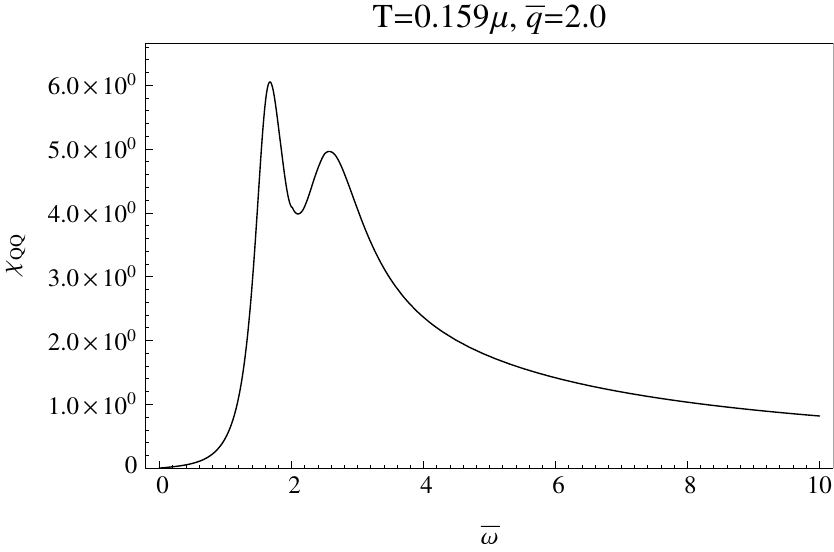}
\includegraphics[scale=0.88]{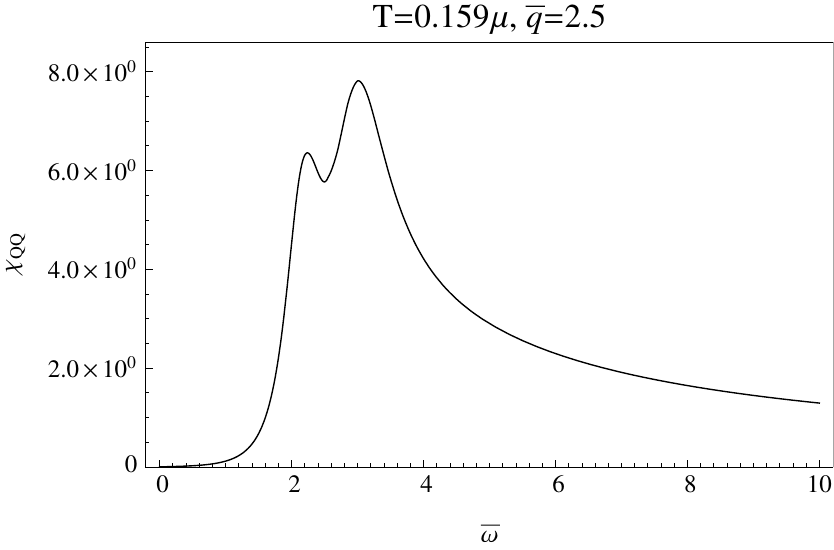}
\includegraphics[scale=0.88]{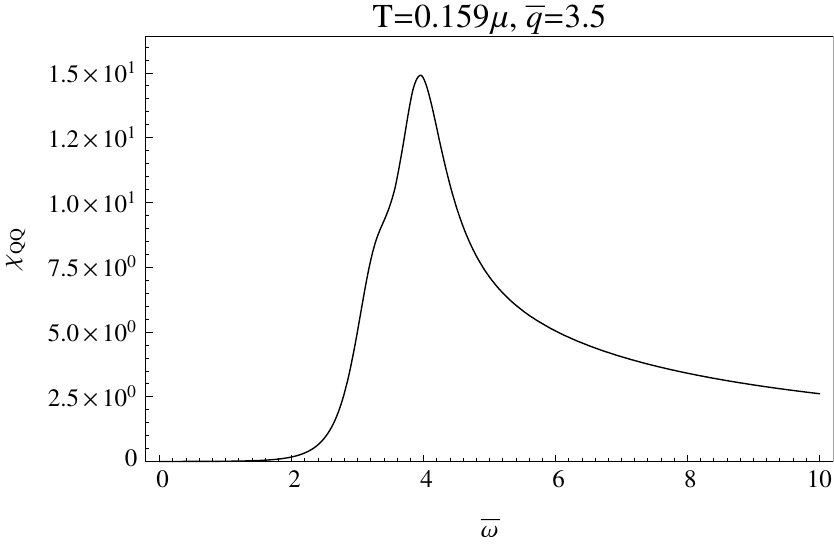}
\includegraphics[scale=0.88]{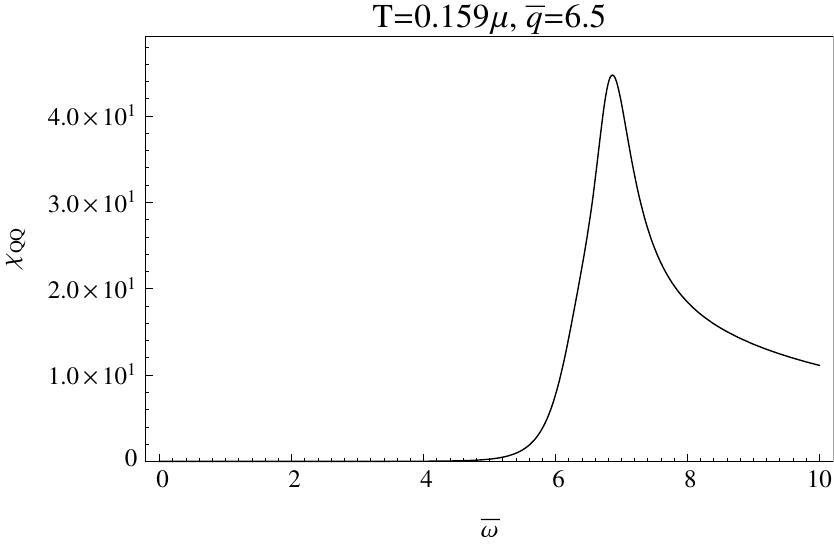}
\caption{The charge density spectral function for $T=0.159\mu$ as the momentum is increased, in units of $2r_0/\kappa_4^2$. As the momentum is increased, the peak due to sound propagation becomes less dominant. An animated version of this figure is available at \url{http://www.physics.ox.ac.uk/users/Davison/RNAdS4animations2.html}.}
\label{fig:qdependencespectralfunctionscharge}
\end{center}
\end{figure*}

\paragraph{}
Finally, we turn our attention to the spectral functions of the theory at low temperatures and as a function of the momentum $\qbar$. These are shown in figures \ref{fig:qdependencespectralfunctionsenergy} and \ref{fig:qdependencespectralfunctionscharge} for $T=0.159\mu$.
At $\qbar=0.5$, the lowest momentum shown, both spectral functions are completely dominated by the peak due to sound propagation. As the momentum is increased, this peak becomes smaller and wider and when $q\gtrsim\mu$, it no longer dominates the spectral function - a peak due to the secondary propagating mode also becomes visible. As the momentum is increased further, these two peaks merge into one peak which moves with speed $\text{Re}\left(\wbar\right)=\qbar$ and constant width. This is a direct reflection of the two corresponding Green's function poles approaching each other in the complex plane. At high momenta, the value of the spectral function is very small at low frequencies $\wbar\lesssim\qbar$ and it only becomes significant when $\wbar\gtrsim\qbar$. Note that there is no significant difference between the charge density and energy density spectral functions in this regime.
\section{An effective hydrodynamic scale}
\label{sec:effectivehydroscale}
\paragraph{}
As discussed previously, when $\mu=0$ there is a long-lived sound mode with momentum $q$ provided that $q\ll T$, and when $T=0$ there is a long-lived sound mode provided that $q\ll\mu$. In the first instance, this is the regime of applicability of hydrodynamics and the condition on the momentum is such that the perturbations occur over much larger distance scales than the mean free path between thermal collisions.

\begin{figure*}
\begin{center}
\subfloat[Numerical results]{\label{fig:soundcontourplot} \includegraphics[scale=0.55]{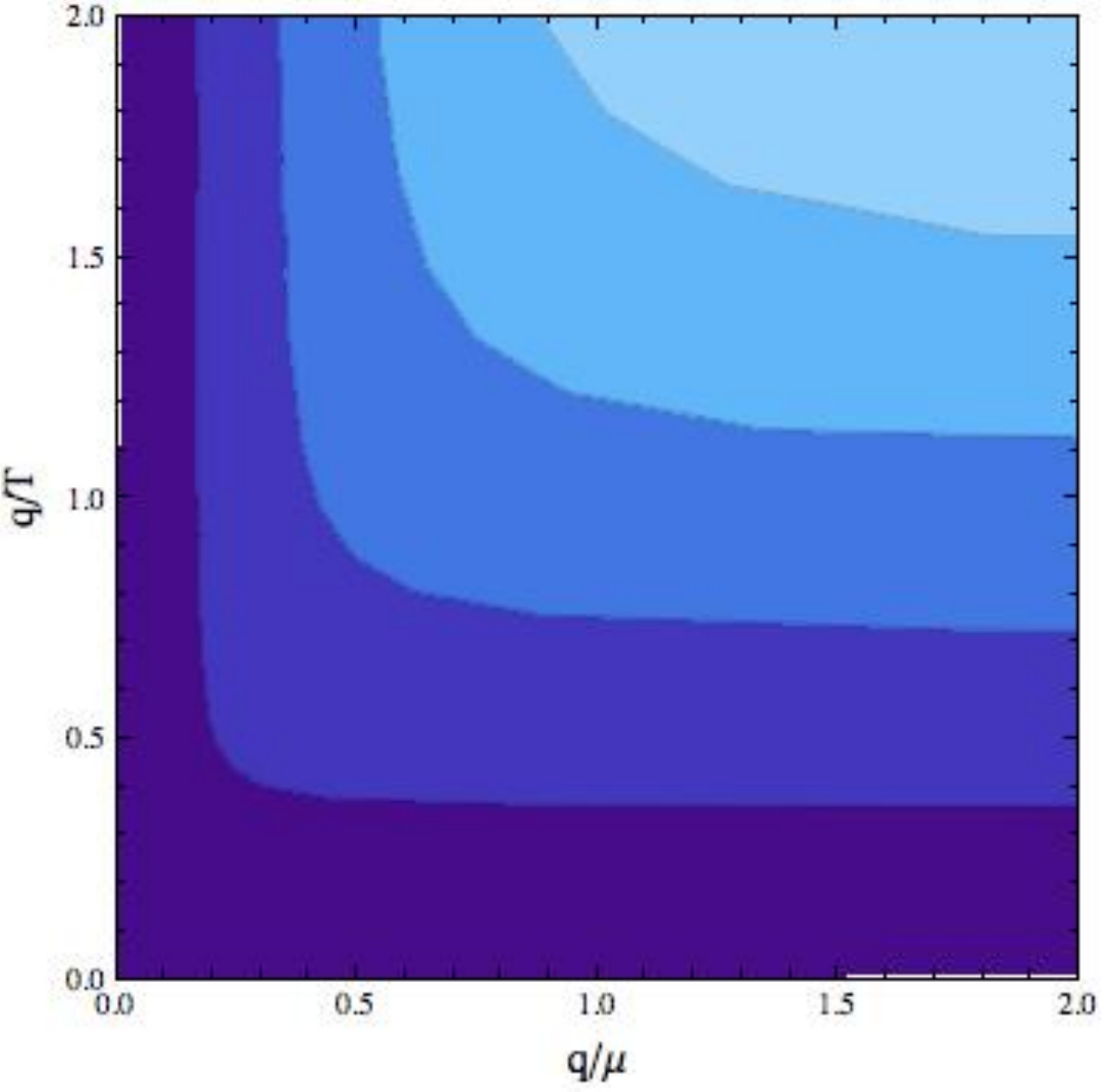}}
\subfloat[Best fit]{\label{fig:soundcontourplotbestfit} \includegraphics[scale=0.55]{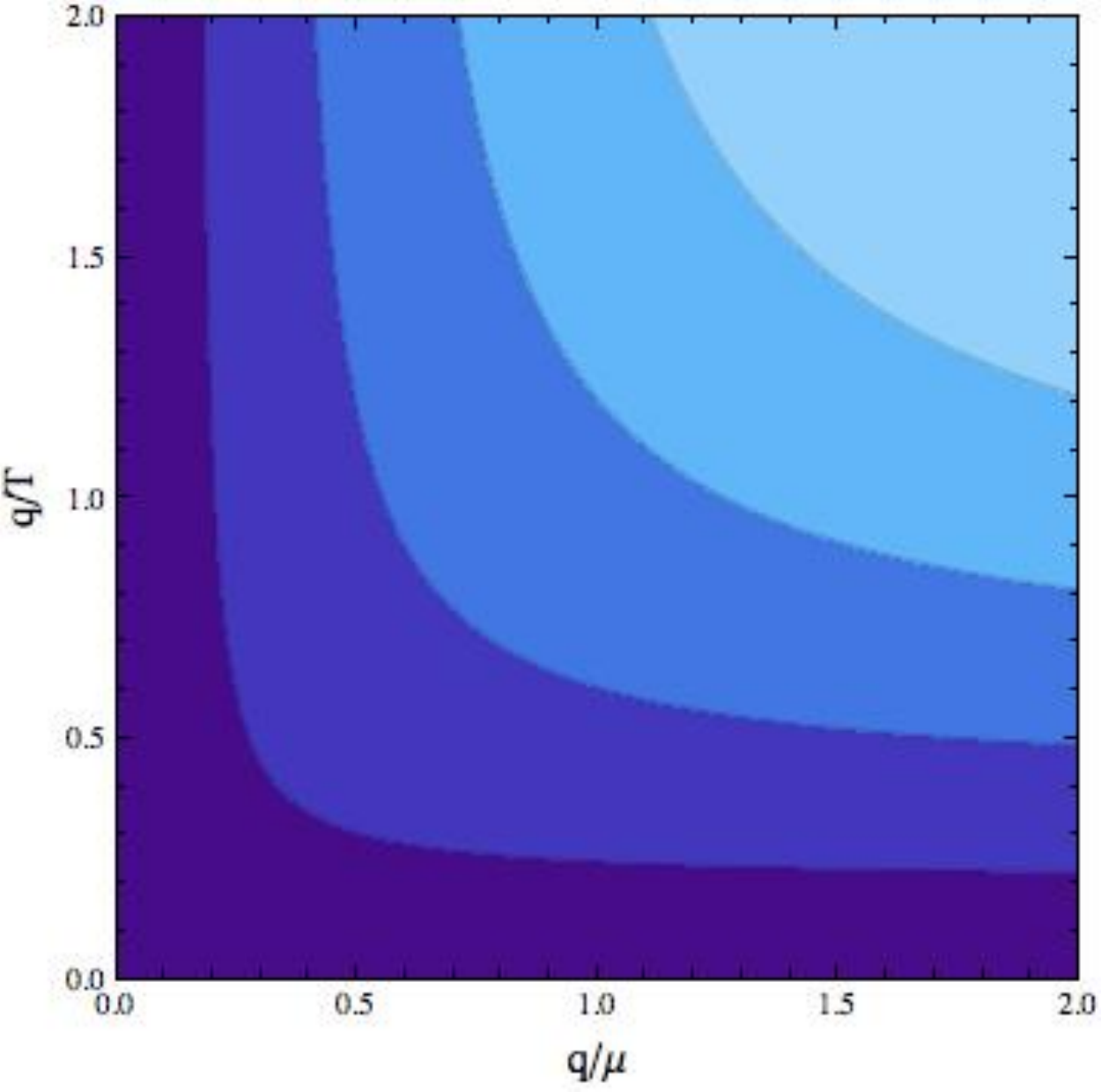}}
\caption{Contour plot showing $|\text{Im}\left(\wbar\right)|/\text{Re}\left(\wbar\right)$ for the sound mode as a function of $q/\mu$ and $q/T$ and the best fit to these results: $a_0=10.1$ and $a_1=8.3$. Darker colours correspond to smaller values (i.e. more stable propagation) and the contours show the values $0.02$, $0.04$, $0.06$, and $0.08$.}
\end{center}
\end{figure*}

\paragraph{}
We have studied the behaviour of the sound mode when both $T$ and $\mu$ are non-zero, to determine if there is some `effective hydrodynamic scale' (or effective mean free path) which determines whether sound propagation is possible or not in this more general case. Figure \ref{fig:soundcontourplot} is a contour plot showing the value of $|\text{Im}\left(\wbar\right)|/\text{Re}\left(\wbar\right)$ - which is the ratio of the decay rate to the propagating frequency - for the sound mode as a function of $q/\mu$ and $q/T$. Darker colours correspond to smaller values (i.e. more stable propagation).
There is a clear pattern in the plot - provided that \textit{one} of $q/T$ or $q/\mu$ is small enough, there is stable sound propagation. It suggests that there is an `effective hydrodynamic scale' governing sound propagation which is qualitatively of the form $E_{\text{eff.}}=T\left(a_0+\ldots+a_{\alpha}\left(\mu/T\right)^\alpha+\ldots+a_1\mu/T\right)$ where $\alpha\in(0,1)$. This reduces to the correct form in the $T=0$ and $\mu=0$ limits separately. A fit of the form $q/(a_0T+a_1\mu)$ could not quantitatively reproduce the plot above - suggesting that this ansatz is an oversimplification (for example, it neglects almost all terms of the form $\left(\mu/T\right)^\alpha$ in the denominator as well as higher order terms in $q$) - but does give qualitatively the correct features for the sound propagation properties. Figure \ref{fig:soundcontourplotbestfit} shows the best fit to this form.

\paragraph{}
An effective hydrodynamic scale of this form is also consistent with the fact that the region $q\ll\mu\ll T$ reproduces the $\mu=0$, $q\ll T$ results, as we saw in section \ref{sec:Tdependencesound}.
\section{Conclusions and discussion}
\label{sec:discussion}

\paragraph{}
In summary, our main results are as follows:

i) When $q\ll\mu$, the long-lived modes of the charge density and energy density Green's functions are the sound and diffusion-like modes with dispersion relations (\ref{eq:sounddispersionrelationallT}) and (\ref{eq:diffusiondispersionrelationallT}) respectively.

ii) When $q\ll\mu$ and $T\ll\mu$, the attenuation of the sound mode shows no significant temperature dependence, unlike in the strongly-coupled D3/D7 field theory and in Landau's theory of Fermi liquids. When $q\ll\mu$ and $T\gg\mu$, the sound and diffusion dispersion relations are well-approximated by the $\mu=0$ results of \cite{Herzog:2003ke,Herzog:2002fn}.

iii) When $q\ll\mu$, the energy density spectral function is dominated by the sound peak at all temperatures, whereas the charge density spectral function undergoes a crossover from sound domination at low temperatures to diffusion domination at high temperatures. This crossover is due to the changing residue at each pole, and occurs at a temperature $T_{\text{cross}}\sim\mu^2/q\gg\mu$.

iv) When $q\gg\mu$ and $T\ll\mu$, the sound and diffusion modes no longer dominate the energy density and charge density spectral functions, and the effects of other modes become important.

v) When both $\mu$ and $T$ are non-zero, a long-lived sound mode will propagate provided that its momentum $q$ is much less than either $\mu$ or $T$.

\paragraph{}
Our results show that although many strongly-coupled field theories at large chemical potential (which have a gravitational dual) possess a $T=0$ sound mode, these are not all LFL-like `zero sound' modes (by which we mean that they don't have the properties shown in figure \ref{fig:LFLplot}) as was the case in the D3/D7 theory. However, there is still the possibility that there could be universal behaviour of the sound mode within subsets of strongly-coupled field theories with a gravitational dual. We note that the density-dependent physics in the D3/D7 and RN-$AdS_4$ theories arise through different holographic mechanisms (see \cite{Hartnoll:2009ns,Charmousis:2010zz} for further discussion of this). In the D3/D7 theory, the background metric is fixed and it is the gauge field action - the DBI action - which alters the equation of motion of the gauge field from the $\mu=0$ Maxwell equation (whose only long-lived mode is the high temperature charge diffusion mode). In contrast to this, the gauge field equation of motion in the RN-$AdS_4$ theory departs from the $\mu=0$ Maxwell equation via couplings to the bulk metric fluctuations. The other major difference between these two field theories is the number of spatial dimensions, but we don't expect that this will have a significant effect on the acoustic properties of the theory (provided that the number of field theory spatial dimensions is greater than one).

\paragraph{}
It would be very interesting to check whether the low temperature sound modes in other probe brane theories \cite{Kulaxizi:2008jx,Hung:2009qk,HoyosBadajoz:2010kd,Lee:2010ez,Bergman:2011rf} share the LFL-like properties of the D3/D7 theory. This would help to establish whether it is the form of the DBI action, which implies that the non-zero density in the field theory is a density of fundamental matter (at least in the cases where the background geometry can be derived from string theory) rather than, for example, the R-charge density in the RN-$AdS_4$ theory, that generates these interesting properties or not. The effects of metric backreaction (i.e. coupling between the field theory's charge density and energy density) are also yet to be computed for these probe brane theories. These may complete the LFL-like picture of acoustic propagation (by reproducing the full LFL sound attenuation curve - figure \ref{fig:LFLplot} - including the hydrodynamic regime), or they may result in deviations from it. We cannot gain any insight into this from our backreacted RN-$AdS_4$ results because of the difference in gauge field actions described previously. Although an expansion of the DBI action in powers of $F_{\mu\nu}$ yields the Maxwell action at lowest order, in field theory quantities this is an expansion in powers of $\mu/T$ which is the opposite limit from that in which any LFL-like properties of a theory would be exhibited.

\paragraph{}
In addition to those mentioned above, there are numerous other field theories with a gravitational dual which may possess interesting sound modes when $T\ll\mu$. Among the most interesting of these are dilatonic black holes \cite{Gubser:2009qt,Goldstein:2009cv,Charmousis:2010zz} and geometries where the bulk charge density is sourced by fermions \cite{Hartnoll:2010gu,Hartnoll:2010ik,Puletti:2010de,Iqbal:2011in,Sachdev:2011ze}. It would also be worthwhile to determine the acoustic properties of more general truncations of supergravity which admit more complicated solutions than RN-$AdS_4$ (for example, those of \cite{Duff:1999gh,Cvetic:1999xp,Gubser:2009qt}), to determine whether the specific truncation chosen has a significant effect on these properties.

\paragraph{}
We have seen that even this relatively simple holographic theory has many non-trivial features in its bosonic excitations. Among the most intriguing are the accuracy of the `zero temperature hydrodynamics', and the crossover temperature $T_{\text{cross.}}\sim\mu^2/q$ between the charge density spectral function being dominated by the sound mode and the diffusion mode. It would be useful to have a clearer physical understanding of these properties, and also to determine if they are present in other field theories at non-zero chemical potential.

\acknowledgments{
We would like to thank Andrei Starinets for very helpful discussions and encouragement, as well as for comments on a draft of this manuscript. We are also grateful to Andy O'Bannon for useful discussions and to Mohammad Edalati for correspondence. R.~A.~D.~ is supported by a studentship from the UK Science and Technology Facilities Council. N.~K.~K.~ is supported by a scholarship from the Greek State Scholarships Foundation.}

\appendix
\section{Equations of motion and action}
\label{sec:appendix}
\paragraph{}
This appendix contains the equations of motion and action used to compute our $T>0$ numerical results. Our $T=0$ results were obtained in the way described in \cite{Edalati:2010pn}.

\subsection{Equations of motion}

\paragraph{}
There are two linearly-independent, coupled equations of motion for $\bar{Z}_1\left(u,\wbar,\qbar\right)$ and $\bar{Z}_2\left(u,\wbar,\qbar\right)$ which can be written in the form
\begin{equation}
\begin{aligned}
&\bar{Z}_1''(u)+A_1\bar{Z}_1'(u)+A_2\bar{Z}_2'(u)+A_3\bar{Z}_1(u)+A_4\bar{Z}_2(u)=0,\\
&\bar{Z}_2''(u)+B_1\bar{Z}_1'(u)+B_2\bar{Z}_2'(u)+B_3\bar{Z}_1(u)+B_4\bar{Z}_2(u)=0,
\end{aligned}
\end{equation}
where we have suppressed the dependence of $\bar{Z}_{1,2}$ on $\wbar$ and $\qbar$, and the coefficients are
\begin{equation}
\begin{aligned}
&A_1=\frac{u^5f'(u)\wbar^2+2f(u)\left[Q^2\qbar^2+u^4\left(\wbar^2-f(u)\qbar^2\right)\right]}{u^5f(u)\left(\wbar^2-f(u)\qbar^2\right)},\\
&A_2=-i\qbar\frac{u^5f'(u)\qbar^2+2\left[Q^2\qbar^2-u^4\left(\wbar^2-f(u)\qbar^2\right)\right]}{u^6\left(\wbar^2-f(u)\qbar^2\right)\left[uf'(u)\qbar^2-4\left(\wbar^2-f(u)\qbar^2\right)\right]},\\
&A_3=\frac{Q^2\left[u^2\left(\wbar^2-f(u)\qbar^2\right)-4f(u)\right]}{u^6f(u)^2}+\frac{4Q^2\qbar^2}{u^6\left(\wbar^2-f(u)\qbar^2\right)}\\
&\hspace{10 mm}+\frac{8Q^2\qbar^2\left[u^4\left(\wbar^2-f(u)\qbar^2\right)+Q^2\qbar^2\right]}{u^{10}\left(\wbar^2-f(u)\qbar^2\right)\left[uf'(u)\qbar^2-4\left(\wbar^2-f(u)\qbar^2\right)\right]},\\
&A_4=i\qbar\frac{4Q^2+u^5f'(u)}{u^7f(u)\left[uf'(u)\qbar^2-4\left(\wbar^2-f(u)\qbar^2\right)\right]},\\
&B_1=-i\qbar\frac{2Q^2\left[uf'(u)\qbar^2-4\left(\wbar^2-f(u)\qbar^2\right)\right]}{u^4\left(\wbar^2-f(u)\qbar^2\right)},\\
&B_2=\frac{1}{u^5f(u)\left(\wbar^2-f(u)\qbar^2\right)\left[uf'(u)\qbar^2-4\left(\wbar^2-f(u)\qbar^2\right)\right]}\Biggl\{-16u^4f(u)^3\qbar^4   \\
&\hspace{10 mm}+2f(u)^2\qbar^2\left[-4Q^2\qbar^2+16u^4\wbar^2+u^5\left(f'(u)+uf''(u)\right)\qbar^2\right]\\
&\hspace{10 mm}-f(u)\Bigl[-8Q^2\qbar^2\wbar^2+16u^4\wbar^4+u\qbar^2\Bigl\{f'(u)\Bigl(2Q^2\qbar^2-2u^4\wbar^2+u^5f'(u)\qbar^2\Bigr)\\
&\hspace{10 mm}+2u^5f''(u)\wbar^2\Bigr\}\Bigr]+u^5f'(u)\wbar^2\left[uf'(u)\qbar^2-4\wbar^2\right]     \Biggr\},\\
&B_3=\frac{8iQ^2\qbar}{u^9f(u)\left(\wbar^2-f(u)\qbar^2\right)\left[uf'(u)\qbar^2-4\left(\wbar^2-f(u)\qbar^2\right)\right]}\Biggl\{u^5f'(u)\wbar^2[uf'(u)\qbar^2-4\wbar^2]\\
&\hspace{10 mm}-f(u)\qbar^2\left[-4Q^2\wbar^2+uf'(u)\left(Q^2\qbar^2-3u^4\wbar^2+u^5f'(u)\qbar^2\right)+u^6f''(u)\wbar^2\right]\\
&\hspace{10 mm}+f(u)^2\qbar^4\left[-4Q^2+u^5\left(f'(u)+uf''(u)\right)\right]\Biggr\},\\
&B_4=\frac{1}{u^4f(u)^2\left[uf'(u)\qbar^2-4\left(\wbar^2-f(u)\qbar^2\right)\right]}\Biggl\{-4Q^2f(u)^2\qbar^4+Q^2\wbar^2\left[uf'(u)\qbar^2-4\wbar^2\right]\\
&\hspace{10 mm}+f(u)\qbar^2\left[8Q^2\wbar^2+uf'(u)\left(u^4f''(u)+5u^3f'(u)-Q^2\qbar^2\right)\right]\Biggr\}.\\
\end{aligned}
\end{equation}
\subsection{Action}
\paragraph{}
In these variables, the off-shell action to quadratic order in the fluctuations is of the form
\begin{equation}
\begin{aligned}
S=\frac{r_0}{2\kappa_4^2}\int_1^\infty du\frac{d\omega d^2q}{\left(2\pi\right)^3}&\Bigl[\mathcal{G}_{11}\partial_u\bar{Z}_1(u,-\wbar,-\qbar)\partial_u\bar{Z}_1(u,\wbar,\qbar)+\mathcal{G}_{12}\partial_u\bar{Z}_1(u,-\wbar,-\qbar)\partial_u\bar{Z}_2(u,\wbar,\qbar)\\
&+\mathcal{G}_{21}\partial_u\bar{Z}_2(u,-\wbar,-\qbar)\partial_u\bar{Z}_1(u,\wbar,\qbar)+\mathcal{G}_{22}\partial_u\bar{Z}_2(u,-\wbar,-\qbar)\partial_u\bar{Z}_2(u,\wbar,\qbar)+\ldots\Bigr],
\end{aligned}
\end{equation}
where the coefficients are 
\begin{equation}
\begin{aligned}
&\mathcal{G}_{11}=\frac{2u^2f(u)}{\wbar^2-f(u)\qbar^2},\\
&\mathcal{G}_{12}=-\frac{2i\qbar uf(u)}{\left(\wbar^2-f(u)\qbar^2\right)\left[-4\left(\wbar^2-f(u)\qbar^2\right)+uf'(u)\qbar^2\right]},\\
&\mathcal{G}_{21}=-\mathcal{G}_{12},\\
&\mathcal{G}_{22}=\frac{2u^4f(u)\left[\left(\wbar^2-f(u)\qbar^2\right)+\frac{\qbar^2Q^2}{u^4}\right]}{Q^2\left(\wbar^2-f(u)\qbar^2\right)\left[-4\left(\wbar^2-f(u)\qbar^2\right)+\qbar^2uf'(u)\right]^2},\\
\end{aligned}
\end{equation}
and the `$\ldots$' represents terms with less than two $u$ derivatives (which cannot generically be written in terms of these gauge-invariant variables). These coefficients, combined with the equations of motion listed previously, allow us to compute the Green's function's poles and spectral functions $\chi_{\epsilon\epsilon}$, $\chi_{QQ}$ by following the method of \cite{Kaminski:2009dh}. Note that the counterterms in (\ref{eq:einsteinmaxwellaction}) (listed, for example, in \cite{Edalati:2010pn}) do not affect these quantities.

\paragraph{}
The numerical check described in \cite{Kaminski:2009dh} relies on the coefficients of the one-derivative terms in the action in addition to the two-derivative terms. Hence to obtain a numerical check in our gauge-invariant formalism, we added a boundary `counterterm' (distinct from those mentioned previously) to the off-shell action of the form
\begin{equation}
S^{\text{c.t.}}=\frac{r_0}{2\kappa_4^2}\int du\frac{d\omega d^2q}{\left(2\pi\right)^3}\frac{d}{du}\left[\phi_I\left(u,-\wbar,-\qbar\right)\mathcal{D}_{IJ}^{\text{c.t.}}\left(u,\wbar,\qbar\right)\phi_J\left(u,\wbar,\qbar\right)\right],
\end{equation}
where $\mathcal{D}_{IJ}^{\text{c.t.}}\left(u,\wbar,\qbar\right)$ was chosen such that the full off-shell action could then be written in terms of the gauge-invariant variables $\bar{Z}_{1,2}$. This boundary term does not alter the equations of motion and as the coefficients $\mathcal{D}_{IJ}^{\text{c.t.}}$ are purely real, it does not have any effect upon the poles of the Green's functions or the diagonal spectral functions $\chi_{\epsilon\epsilon}$ and $\chi_{QQ}$. We do not list the coefficients here as they are very lengthy and do not directly affect the results presented. For our numerical computations, we found it more convenient to use the radial co-ordinate $z\equiv1/u$. 

\bibliographystyle{JHEP}
\bibliography{RNAdS4FinalDraft}

\end{document}